\documentclass[default]{svmult}

\usepackage{type1cm}        
\usepackage{makeidx}         
\usepackage{graphicx}        
\usepackage{multicol}        
\usepackage[bottom]{footmisc}

\usepackage{newtxtext}       %
\usepackage[varvw]{newtxmath}       

\usepackage{comment}
\usepackage{aas_macros}     
\usepackage{wrapfig}

\usepackage[breaklinks,colorlinks,urlcolor=blue,citecolor=blue,linkcolor=magenta]{hyperref}

\usepackage{enumitem}

\newcommand{\be}{\begin{equation}} 
\newcommand{\ee}{\end{equation}}
\newcommand{\bea}{\begin{equation}\begin{aligned}} 
\newcommand{\eea}{\end{aligned}\end{equation}}

\def\lsim{\mathrel{\raise.3ex\hbox{$<$\kern-.75em\lower1ex\hbox{$\sim$}}}}
\def\gsim{\mathrel{\raise.3ex\hbox{$>$\kern-.75em\lower1ex\hbox{$\sim$}}}}

\newcommand{\tr}{{\rm tr}}
\newcommand{\yr}{{\rm yr}}
\newcommand{\kyr}{{\rm kyr}}

\newcommand{\Gpc}{{\rm Gpc}}
\newcommand{\tH}{t_0}

\newcommand{\td}{{\rm d}}
\newcommand{\Msun}{M_\odot}

\makeindex             

\begin{document}


\graphicspath{{Figures/}}

\title*{Formation of primordial black hole binaries and their merger rates}
\author{Martti Raidal, Ville Vaskonen and Hardi Veerm\"{a}e}
\institute{
Martti Raidal \at 
Keemilise ja Bioloogilise F\"u\"usika Instituut, R\"avala pst. 10, 10143 Tallinn, Estonia \\ 
\email{martti.raidal@cern.ch} 
\and Ville Vaskonen \at 
Keemilise ja Bioloogilise F\"u\"usika Instituut, R\"avala pst. 10, 10143 Tallinn, Estonia, \\
Dipartimento di Fisica e Astronomia, Universit\`a degli Studi di Padova, Via Marzolo 8, 35131 Padova, Italy \\
Istituto Nazionale di Fisica Nucleare, Sezione di Padova, Via Marzolo 8, 35131 Padova, Italy, \\
\email{ville.vaskonen@pd.infn.it}
\and Hardi Veerm\"{a}e \at 
Keemilise ja Bioloogilise F\"u\"usika Instituut, R\"avala pst. 10, 10143 Tallinn, Estonia \\ 
\email{hardi.veermae@cern.ch}}

\maketitle\label{chapter:GWmergers}


\abstract{We review the theory behind the formation of primordial black hole binaries and their merger rates. We consider the binary formation in the early and late Universe, emphasising the former as it gives the dominant contribution of the present primordial black hole merger rate. The binaries formed in the early Universe are highly eccentric so their interactions with other primordial black holes can significantly increase their coalescence times and thereby suppress the merger rate. We discuss in detail how the suppression of the merger rate arising from such interactions can be estimated and how such interactions lead to the formation of another, much harder, binary population that contributes to the present merger rate if more than 10\% of dark matter consists of primordial black holes with a relatively narrow mass distribution. When the primordial abundance is below 1\%, encounters between primordial black holes are unlikely and their effect on the merger rate becomes negligible.  }

\newpage

\section*{Content}
\setcounter{minitocdepth}{2}
\dominitoc


\section{Introduction}\label{sec:1}

Gravitational wave (GW) signals created by primordial black hole (PBH) binaries provide a direct probe of hypothetical PBH populations. The PBH populations are expected to differ from astrophysical black hole (ABH) populations in aspects such as mass distribution, redshift dependence of the merger rate, spins or even spatial distribution. A good theoretical understanding of both populations is crucial for employing the full potential of GW observations. In the following, we will focus on describing the PBH binary populations and the resulting PBH merger rates. 

The characteristics of the present-day PBH populations are moderately model-dependent. They can depend on the numerous ways how the PBHs could have been produced in the early Universe and potentially also on deviations from the standard cosmology, which can affect the evolution of the PBH populations after their formation. To our luck, it is not necessary to specify the details of PBH formation to estimate the properties of the present-day PBH binary populations. Only the PBH abundance, the distribution of their masses and spins as well as their spatial distribution before matter-radiation equality should be specified. This sets the initial stage of PBH structure formation as nearby PBH pairs decouple from the expansion and form the first bound structures in the Universe.

In certain scenarios, a population of merging PBH binaries can be probed by GW experiments. With enough data, these mergers can help us reconstruct the initial characteristics of PBHs in the early Universe and begin probing the early Universe by reconstructing scenarios that would produce such initial conditions. The goal of this chapter is to outline the first part of this exercise and show how an initial population of PBHs coupled to the Hubble flow in the very early Universe will give rise to a population of PBH binaries merging in the present day.

\subsection{Binary formation channels}
\label{sec:channel_summary}

There are various channels for forming PBH binaries. These can be classified by 1)~whether the binaries were formed in the early or late Universe, and 2)~whether the formation included dynamics of two or more bodies. The PBH binary characteristics of each combination can be outlined as follows:
\begin{enumerate}
    \item In the {\bf early two-body channel}, binaries are formed from nearby PBH pairs that decouple the Hubble flow. Such binaries generally dominate the PBH binary merger rate when $f_{\rm PBH}$ - the fraction of DM in PBHs - is small. However, they are very eccentric since most of their angular momentum comes from tiny tidal forces induced by fluctuations in the density of surrounding matter. This makes the merger rate susceptible to perturbations by encounters with other PBHs, so their subsequent evolution should be accounted for when estimating their present merger rate. In particular, when $f_{\rm PBH} \gtrsim 0.1$, small-scale structure formation is enhanced. This leads to greater disruption of these binaries and suppresses the corresponding merger rate.

    \vspace{1mm}
    \item In the {\bf early three-body channel}, bound three-body systems are formed after initial compact 3-PBH configurations decouple from the Hubble flow. In such a system, a binary remains if one of the PBHs is ejected or if two of the PBHs merge. The resulting binaries are much less eccentric and, therefore, much harder~\footnote{A binary is considered hard when its binding energy is larger than the average kinetic energy of surrounding PBHs.} than the binaries from the early two-body channel. As a result, their merger rate is not significantly affected by encounters with other PBHs. However, since compact 3-PBH systems are less likely to occur in the early Universe, this channel is generally subdominant to the early two-body channel for small $f_{\rm PBH}$. However, it can dominate if $f_{\rm PBH}\approx 1$ or if the PBHs are formed in clusters.

    \vspace{1mm}
    \item In the {\bf late two-body channel}, PBH binaries are formed from hyperbolic encounters of PBHs in DM haloes if they lose a sufficient amount of energy via GW emission. The resulting binaries are typically eccentric and merge relatively fast. Although the resulting merger rate depends on the PBH structure formation and may be boosted by spiky DM profiles or PBH clustering, such binaries tend to have a smaller merger rate than PBH binaries formed in the early Universe. Although this channel could also produce ABH-PBH binaries, their contribution is expected to be marginal~\cite{Kritos:2020wcl}.

    \vspace{1mm}
    \item The {\bf late three-body channel} can be active in the environment with large PBH densities. The binaries are formed in 2+1 PBH encounters, where the 3rd PBH carries away enough energy for the remaining 2 to become bound. Analogously to the early three-body channel, these binaries are hard and not too eccentric. Although their formation implies an environment with relatively frequent PBH-PBH binary encounters, such encounters will not significantly affect their merger rate. This channel is generally weaker than the ones listed above. It can be enhanced if $f_{\rm PBH}\approx 1$ or if the formation of PBH structures is enhanced, i.e., by initial clustering.
\end{enumerate}
In the following, we will consider each channel in detail.

\subsection{Preliminaries}
\label{sec:preliminaries}

We will use geometric units $c=G=1$ throughout this chapter. We consider a generic PBH mass function, that we define as
\be
	\psi(m) \equiv \frac{m}{\rho_{\rm PBH}} \frac{\td n_{\rm PBH}}{\td \ln m} \,,
\ee 
where $\rho_{\rm PBH}$ and $n_{\rm PBH}$ denote the total energy and comoving number densities of PBHs. With this definition we get $\int \psi(m) \td \ln m = 1$ and the average of a quantity $X$ is given by
\be
    \langle X \rangle  
    \equiv \int \frac{\td n_{\rm PBH}}{n_{\rm PBH}} \, X  
    = \int \td \ln m \, \frac{\langle m\rangle}{m} \,\psi(m) X  \,,
\ee
where $\langle m\rangle = \rho_{\rm PBH}/n_{\rm PBH} = 1/\int \td \ln m \, m^{-1} \,\psi(m)$. 

We denote the masses of individual PBHs by a lowercase $m$ and the PBHs making up the binary by $m_1$ and $m_2$. The total mass and the reduced mass of the binary are $M \equiv m_1 + m_2$ and $\mu \equiv m_1m_2/M$ and we often use the symmetric mass ratio $\eta \equiv \mu/M$ to quantify unequal mass binaries.

We work in the Newtonian limit. Thus, the binding energy of the binary is
\be\label{eq:2body_E}
    E_{\rm bin} = \frac{\mu \dot{\mathbf{r}}^2}{2}-\frac{M \mu}{r} = \frac{M \mu }{2 r_a},
\ee
where $r_a$ denotes the semimajor axis of the binary, $\mathbf{r} = \mathbf{r}_1 - \mathbf{r}_2$ is the separation of the PBHs positioned at $\mathbf{r}_{1}$ and $\mathbf{r}_{2}$, and the dot denotes time derivation. The binary's angular momentum ${\bf L} = \mu \mathbf{r} \times \dot{\mathbf{r}}$ will be quantified via the dimensionless quantity
\be\label{eq:2body_j}
	{\bf j} \equiv \frac{{\bf L}/\mu}{\sqrt{r_a M}} \,,
\ee
which satisfies $j\equiv |{\bf j}| \leq 1$ and is related to the eccentricity as
\be\label{eq:2body_e}
    e = \sqrt{1-j^2}.
\ee
Thus, for our purposes, it is sufficient to characterize PBH binaries by four quantities, e.g., $m_1$, $m_2$, $r_a$ and $j$. 

To estimate the merger rate, we need to consider the formation of PBH binaries and their disruption between formation and merger. Given the PBH binary formation rate density $\td R_{\rm b}(t')/(\td m_1 \td m_2 \td j \td r_a)$ at time $t'$, the merger rate density at time $t$ is
\be\label{eq:Rprel}
    \frac{\td R(t)}{\td m_1 \td m_2} 
    = \int \!\td t' \td j \td r_a \, \frac{\td R_{\rm b}(t')}{\td m_1 \td m_2 \td j \td r_a} \delta\left(t - t' - \tau(m_1,m_2,r_a,j)\right)\, ,
\ee
where $\tau$ denotes the coalescence time. The coalescence time can be estimated assuming the binary will evolve only via GW emission after its formation. In the Newtonian regime, the coalescence time of eccentric binaries is~\cite{Peters:1964zz}
\be\label{eq:tau}
    \tau = \frac{3}{85}\frac{r_a^4 j^7}{\eta M^3} \,.
\ee
This expression would overestimate the coalescence time mostly by a factor of 1.81 when applied to almost circular binaries.\footnote{For any $j$, the time evolution of $j$ and $r_a$ of purely GW driven binaries~\cite{Peters:1964zz} can be integrated to give,
\be\label{eq:tau_gen}
    \tau 
    = \frac{3 r_a^4 j^7}{85 \eta M^3}
    \frac{F(j)-F(1) j}{ \left(1-\frac{121}{425}j^2\right)^{3480/2299} \left(1-j^2\right)^{24/19}}
    \approx \frac{3 r_a^4 j^7}{85 \eta M^3} \frac{1 - 0.70 j + 0.62 j^2}{1 + 0.67 j}\,,
\ee
where $F(j) \equiv F_1\left(-1/2;-1181/2299,-5/19;1/2;j^2 121/425,j^2\right)$ with $F_1$ the first Appell hypergeometric function. The second approximation deviated from the exact one by less than 0.5\%. The limiting case of circular binaries, $j \to 1$, gives $\tau = (5/256) r_a^4/\eta M^3$.} If the binary is disrupted\footnote{Disruption in this context means any event that leads to a change in the orbital parameters that significantly affects the coalescence time, ionization or swapping out a constituent of the binary.}, or evolves between formation and merger due to effects other than GW emission, e.g., accretion, then \eqref{eq:Rprel} must be modified accordingly.

\newpage
\section{PBH binary formation in the early Universe}
\label{sec:earlyPBBH}

In the early Universe, PBH binaries can form when nearby PBHs decouple from the Hubble flow~\cite{Nakamura:1997sm}. The formation process and the resulting merger rate depend on whether a third PBH falls into the pair. In this section, we will first lay out the derivation for the two-body formation channel, and look at phenomena that may perturb these binaries and suppress this rate. Finally, we will consider the merger rate stemming from the three-body channel.

\subsection{Dynamics of early binary formation}
\label{sec:dynamics}

\begin{figure}[t]
\begin{center}
\includegraphics[width=0.98\textwidth]{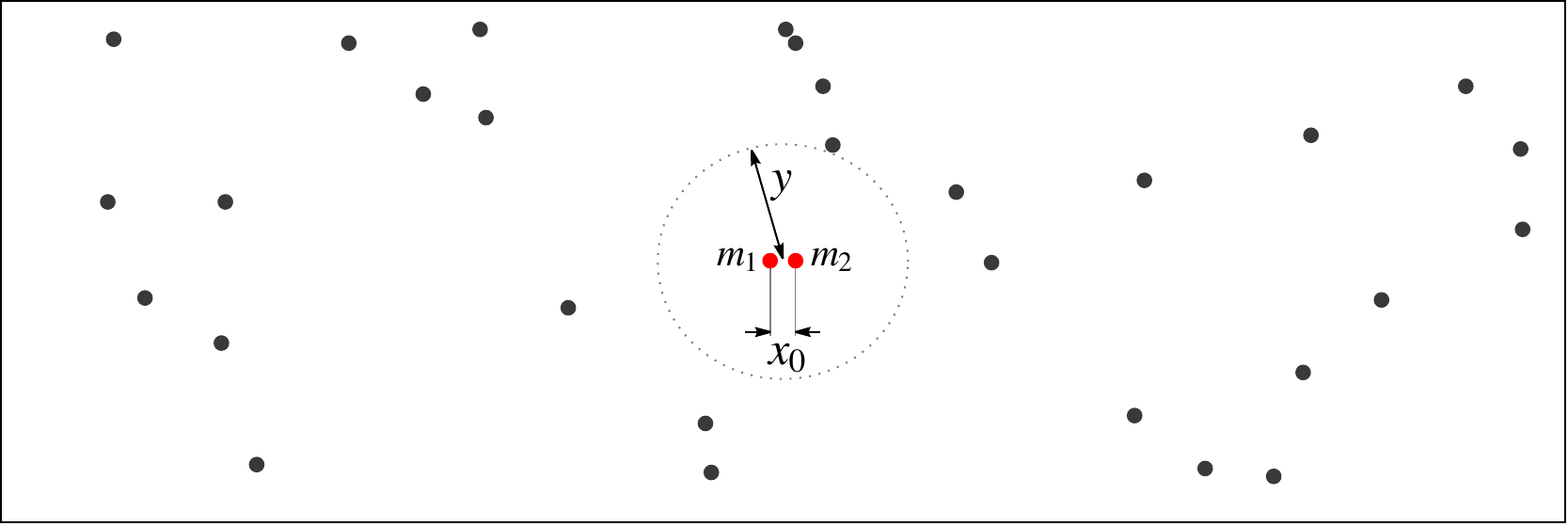}
\caption{A initial PBH configuration and its parametrization. }
\label{fig:setup}
\end{center}
\end{figure}

To set up the initial configuration for the early Universe two-body channel, let us consider a pair of PBHs with masses $m_1$ and $m_2$ at a comoving separation $x_{0}$ which initially follow the Hubble flow. We assume that there are no PBHs within a comoving radius $y$ of the pair, to guarantee that the pair evolves as a two-body system after decoupling. This initial setup is shown in Fig.~\ref{fig:setup}. The size of the empty region will be determined later by requiring that the PBHs closest to the pair will not become gravitationally bound to the pair after decoupling from the Hubble flow and, therefore, form a bound system consisting of 3 or more bodies. Such systems will be addressed later in Sec.~\ref{sec:RE3}. We further assume that the positions of PBHs are uncorrelated, that is, they follow a Poisson distribution. In this case, the comoving number density of PBH configurations described above is
\bea\label{eq:n_b}
	\td n_{\rm pairs}
&	=	\frac{1}{2} e^{-\bar{N}(y)} \td n(m_1) \td n(m_2) \td V(x_{0}) \,,
\eea
where $\td n(m)$ is the comoving number density of PBHs in the mass range $(m, m+ \td m)$, and 
\be
    \bar{N}(y) \equiv n V(y) 
\ee
is the expected number of PBHs in a spherical volume $V(y) \equiv (4\pi/3) y^3$ with a comoving radius $y$. The factor $1/2$ avoids overcounting.

The population of PBH binaries created from such initial conditions is described by the formation rate
\be
    \frac{\td R_{\rm b}(t)}{\td m_1 \td m_2 \td j \td r_a} 
    = \int \td x_0 \frac{\td n_{\rm p}}{\td m_1 \td m_2 \td x_0} \frac{\td P (j, r_a| x_0, y)}{\td j \td r_a} \delta(t - t_{\rm dc}(x_0)) \,,
\ee
where $t_{\rm dc}(x_0)$ denotes the time of decoupling of the pair from the Hubble flow and $\td P (j, r_a| x_0,y)/\td j \td r_a$ is the distribution orbital parameters given $x_0$ and $y$.\footnote{We don't show the dependence on the progenitor masses.} A few simplifications can be made. Since the early binaries are formed before the matter-radiation equality, we can omit the dependence on formation time and assume that $t_{\rm dc}(x_0) \approx \tH$, where $\tH$ is the age of the Universe.\footnote{We use $\tH = 13.8 \rm Gyr$ in numerical estimates.~\cite{Planck:2018vyg}} 
Additionally, $P (j, r_a| x_0,y) \propto \delta(r_a - r_a(x_0)) \, \td P (j| x_0,y)/\td j$ since $r_a$ can be predicted from $x_0$ alone. Given the coalescence time~\eqref{eq:tau}, we find that the merger rate density~\eqref{eq:Rprel} can be recast as
\bea\label{dR_early_y}
    \td R^{(y)}_{\rm E2}(t)
&	= \int \td n_{\rm pairs} \td j \frac{\td P(j | x_0,y)}{\td j}  \delta\left(t - \tau(r_a(x_0),j)\right)  \\
&	= \frac{1}{14 \tau} \td n(m_1) \td n(m_2) \int \td V(x_{0}) e^{-\bar{N}(y)}  j \frac{\td P (j | x_0,y)}{\td j} \bigg|_{j = j(\tau, r_a(x_0))} \,,
\eea
where, in the second line, $j(\tau, r_a(x_0))$ is obtained by inverting Eq.~\eqref{eq:tau}. 

This estimate must, however, be modified to account for the disruption of the early binaries via collisions of the binary with other PBHs in DM clusters between the binary formation and merger. We will describe this effect by the survival probability $S_{\rm L}$, which we take to be independent of the initial conditions. Furthermore, the exclusion of initial conditions in which the binary gets disrupted by nearby PBHs shortly after formation can be characterised by another survival probability $S_{\rm E}$. So, the merger rate density Eq.~\eqref{dR_early_y} takes the final form
\bea\label{dR_early}
    \boxed{\td R_{\rm E2}(t) 
    = S_{\rm L} S_{\rm E} \times \td R_{\rm E2}^{(0)}(t)  } \,,
\eea
where $S_{\rm E} \equiv \td R_{\rm E2}^{(y)}/\td R_{\rm E2}^{(0)} \leq 1$.\footnote{Later, the suppression factor will also account for particle DM fluctuations, which are neglected in $R_{\rm E2}^{(0)}$.} Note that $y$ is generally not an independent variable and can depend on various model parameters.

All in all, to proceed, we need to compute (\emph{i}) how the semimajor axis $r_a(x_0)$ is formed during the decoupling from the Hubble flow (see section~\ref{sec:decoupling}), (\emph{ii}) how spatial inhomogeneities set the distribution of initial angular momenta of the binary (see section \ref{sec:jdistribution}), (\emph{iii}) what is the minimal distance $y$ between the pair and other PBHs that would avoid disrupting the binary due to the infall of other close PBHs into the binary and how does it affect $S_{\rm E}$ (see section \ref{sec:RE2}), (\emph{iv}) estimate disruption in late DM haloes by computing the probability $S_{\rm L}$ that the binaries do not collide with other PBHs during before the merger (see section \ref{sec:disruption_haloes}).

\subsubsection{Decoupling}
\label{sec:decoupling}

The line element for spacetimes with small inhomogeneities in the comoving Newtonian gauge and the absence of anisotropic stress is
\be
	\td s^2 = - (1 + 2\phi({\bf x})) \td t^2 +  (1 - 2\phi({\bf x})) a(t)^2 \td x^2 \,,
\ee
where the potential $\phi$ is determined by the total matter energy density $\rho_{\rm M}(\bf x)$ as $\Delta \phi = 4\pi a^2 \rho_M(\bf x)$. Treating the PBHs as point masses, we obtain
\be\label{eq:phi}
	\phi({\bf x}) =
	- \sum_i \frac{m_i}{a |{\bf x}_i - {\bf x}|} 
	- 4\pi \bar\rho_{\rm M} \int \frac{\td^3 k}{(2\pi)^3} \frac{a^2}{k^2} e^{-i{\bf k} \cdot {\bf x} } \delta_{\rm M}({\bf k}),
\ee
where $m_i$ denote the masses and ${\bf x}_i$ the comoving positions of the PBHs, $\bar\rho_{\rm M}$ the average comoving matter energy density and $\delta_{\rm M}(\bf k)$ the matter density perturbation. A system of $N$ non-relativistic PBHs, $a \td x/\td t \ll 1$, is described by the action 
\be\label{action_N}
	S^{(N)} 
	 = \int \td t \bigg[ \sum_{i} m_i \left( \frac{1}{2} \dot {\bf r}_i^2 +  \frac{1}{2} \frac{\ddot a}{a} {\bf r}_i^2 - \phi_{\rm ex}({\bf x})\right) + \sum_{i > j}\frac{m_im_j}{|{\bf r}_i - {\bf r}_{j}|} \bigg] \,,
\ee
where ${\bf r}_{i} \equiv a  {\bf x}_{i}$ denotes the proper distance, the last term accounts for the pairwise interaction of the PBHs and the potential $\phi_{\rm ex}({\bf x})$ describes the effect of the surrounding matter on the $N$-body system. For a PBH pair, the action can be approximated as
\bea\label{action_2}
	 S^{(2)}
	 \approx \int \td t \bigg[ &\frac{M}{2} \left(\dot{\bf r}_c^2 + \frac{\ddot a}{a} {\bf r}_c^2\right) - M \phi_{\rm ex}({\bf r}_c) + \frac{\mu}{2} \left(\dot{\bf r}^2 + \frac{\ddot a}{a} {\bf r}^2 + \frac{2M}{r} - {\bf r}\cdot {\bf T} \cdot  {\bf r} \right)
	\bigg] \,,
\eea
where $\bf r$ denotes the proper separation of the PBH, ${\bf r}_c$ the centre of mass of the PBH pair, 
and ${\bf T}_{ij} \equiv \partial_i\partial_j\phi_{\rm ex}({\bf r}_c)$ is the tidal tensor due to the external forces. In summary, the PBH pair is subject to the following forces
\be\label{eq:forces}
	{\bf F}/\mu
	= \underbrace{{\bf r} \ddot a/a}_{\mbox{Hubble flow}}
	-  \underbrace{M \hat {\bf r}/r^2}_{\mbox{self-gravity}}
	+ \underbrace{(\hat {\bf r} \cdot{\bf T}\cdot{\bf r}) \hat {\bf r}}_{\mbox{radial tidal forces}}
	+ \underbrace{({\bf r} \times ({\bf T}\cdot{\bf r}))}_{\mbox{tidal torque}} \times (\hat {\bf r}/r) \,,
\ee
where $\hat{\bf r} \equiv {\bf r}/r$. The first three forces are radial, whereas the last term provides the torque that prevents the head-on collision of the PBH pair.

The angular momentum ${\bf L}$ of the pair is generated by the tidal torque,
\be\label{eq:2body_L}
    {\bf L} 
    = \int \td  t\, {\bf r} \times {\bf F} 
    = \mu \int \td t\,  {\bf r} \times ({\bf T}\cdot{\bf r}) \,.
\ee
During radiation domination, the PBHs surrounding the pair are assumed to follow the Hubble flow and the matter density perturbations are constant. This implies that the potential~\eqref{eq:phi} scales roughly as $\phi \propto a^{-1}$. Consequently, ${\bf T} \propto a^{-3}$ and the tidal forces get damped fast after the pair decouples from the Hubble flow. The pair has then formed a binary with angular momentum $\bf L$ and semi-major axis $r_a$.

\begin{figure}[t]
\begin{center}
\includegraphics[width=0.98\textwidth]{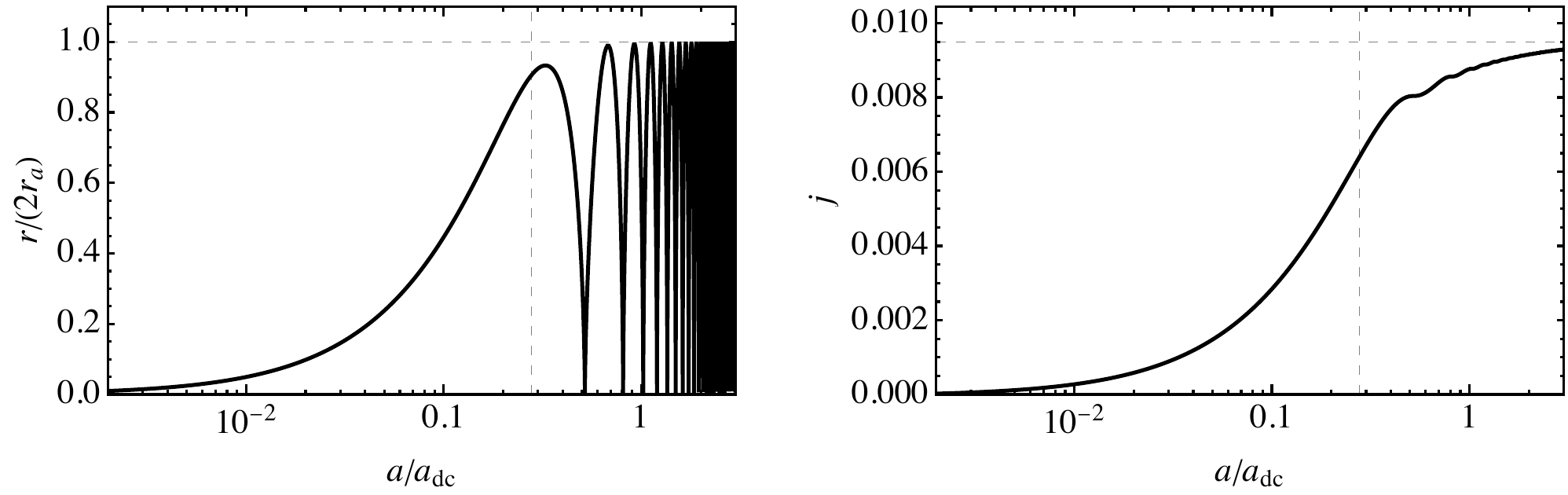}
\caption{Decoupling of the PBH pair. The left and right panels show, respectively, the evolution of the pair's separation and angular momentum in a weak tidal field. The vertical dashed lines show the moment when the self-gravity of the pair starts to dominate over the Hubble flow (see Eq.~\eqref{eq:forces}). The horizontal dashed line in the right panel shows the approximation in Eq.~\eqref{eq:j}.}
\label{fig:decoupling}
\end{center}
\end{figure}

The binary decouples when the Hubble flow becomes subdominant and the second term in \eqref{eq:forces} takes over. To estimate when this happens, consider the overdensity in matter due to the PBH pair of comoving size $x_0$,
\be
        \delta_{\rm pair} \equiv \frac{M/2}{\rho_{\rm M} V(x_0)},
\ee
which should exceed the average matter density $\delta_{\rm pair} \gg 1$ if the PBH pair is to form a binary. Such a region begins to collapse when its density becomes comparable to the density of radiation, $\rho_{\rm R} a^{-4} \approx  \delta_{\rm pair} \rho_{\rm M} a^{-3}$, that is when
\be\label{eq:a_dc}
	a \approx a_{\rm dc} \equiv a_{\rm eq}/\delta_{\rm pair}\, ,
\ee
where $a_{\rm eq}$ denotes the scale factor during matter-radiation equality. Since $\delta_{\rm pair} \gg 1$, we expect the binary to be formed during radiation domination. To make this more precise, consider first the idealized case of a head-on collision for which, the radial equation of motion of the pair, $\ddot{r} - r \ddot a/a + Mr^{-2} = 0$, can be recast in terms of the comoving separation $x \equiv  r/a$ as
\be\label{eom:x_rad}
	x'' + \frac{a}{a_{\rm dc}} \frac{x_0^3}{x^2} = 0 \,,
\ee
where the prime denotes derivation with respect to $\ln(a)$ and the initial conditions are given by $x(a_0) = x_0$, $x'(a_0) = 0$ with $a_0 \ll a_{\rm dc}$. The solution of the equation of motion can be expressed as $x(a) \approx x_0 \chi(a/a_{\rm dc})$, where $\chi(y)$ solves $(y\partial_y)^2\chi+y\chi^{-2} = 0$ with $\chi(y\to 0)=1$, $\chi'(y\to 0)=0$. The function $\chi(y)$ oscillates with an amplitude that decreases asymptotically as $0.2/y$, so the semi-major axis of the binary is
\be\label{eq:r_a}
    r_a = a x_a/2  
    \approx 0.1 a_{\rm dc} x_0 
    \approx 0.84 \rho_{\rm R} x_0^4/M\,.
\ee
The first root of $\chi(y)$ lies at $y \approx 0.54$ corresponding to the first close encounter of the pair at $a \approx 0.54 a_{\rm dc}$ and indicating that the binary decouples from the Hubble flow before that. In Fig.~\ref{fig:decoupling} we indicate by the dashed vertical line the moment when the self-gravity of the pair starts to dominate over the Hubble flow.

The idealised case discussed above is in excellent agreement with the radial evolution in the presence of a weak tidal field illustrated in Fig.~\ref{fig:decoupling} as well as with $N$-body simulations in which the tidal effects are created by dynamical PBHs surrounding the pair~\cite{Raidal:2018bbj} suggesting that the angular momentum can be studied perturbatively as long as the orbit remains eccentric ($j\ll 1$)~\cite{Ali-Haimoud:2017rtz}. Plugging the solution of \eqref{eom:x_rad} into \eqref{eq:2body_L}, we find that the dimensionless angular momentum \eqref{eq:2body_j} is 
\bea\label{eq:j}
	{\bf j} 
&	= \frac{1}{\sqrt{r_{a} M}}\int \frac{\td a}{a^{2}H}\, {\bf x} \times ({\bf T}a^{3} \cdot {\bf x})
	= \frac{0.95 \, x_0^3}{M} \, \hat {\bf r} \times ({\bf T}a^3\cdot \hat {\bf r}) \,,
\eea
where we used $\int^{\infty}_{0} \td y\,  \chi(y)^2 \approx 0.3$. Note that under the above assumptions ${\bf T}a^3$ is constant. 

\subsubsection{Distribution of angular momenta}
\label{sec:jdistribution}

The tidal torque is determined by the distribution of surrounding PBHs as well as the inhomogeneities of the rest of matter. The resulting dimensionless angular momentum can thus be expressed as
\be \label{eq:j_total}
    {\bf  j} = {\bf  j}_{\rm PBH} + {\bf  j}_{\rm M},   \qquad
    {\bf  j}_{\rm PBH} \equiv \sum_i {\bf  j}_1({\bf x}_i, m_i)\, ,
\ee
where ${\bf  j}_1$ is the angular momentum generated by single PBH with mass $m_i$ at a comoving distance $x_i$ from the pair, and ${\bf  j}_{\rm M}$ is the angular momentum generated by matter fluctuations. Using Eq.~\eqref{eq:j} with ${\bf T}_{i} =  m_i ({\bf 1} - 3\hat {\bf x}_i \otimes \hat {\bf x}_i )/x_i^3 a^{-3}$ and defining 
\bea\label{def:j0}
	j_0 \equiv  0.95 \, \bar{N}(x_{0}) \frac{\langle m \rangle}{M} \approx  0.4 \frac{f_{\rm PBH}}{\delta_{\rm pair}} \,,
\eea
where $\langle m \rangle = \rho_{\rm PBH}/n$ is the average PBH mass and we have used $\Omega_{\rm M}/\Omega_{\rm DM} = 1.2$, we get
\be\label{j1}
	{\bf  j}_1({\bf x}_i, m_i) 
	=    j_0 \frac{m_i}{\langle m \rangle} \frac{3}{\bar{N}(x_i)}  \hat{\bf x}_i \times  \hat {\bf r}  \,(\hat {\bf x}_i \cdot \hat {\bf r}) \,.
\ee 
The quantity $j_0$ is a reference scale for the dimensionless angular momentum. It captures its order of magnitude because torque is dominantly generated by the closest PBHs, for which $\bar{N}(x_{i}) \approx 1$, and the remaining terms in Eq.~\eqref{j1} will contribute an $\mathcal{O}(1)$ on average. Binaries form from PBH pairs which satisfy $\delta_{\rm pair} \gg 1$, thus we naturally expect that $j_0 \ll 1$.

To compute the distribution of angular momenta, we will consider the most common scenario in which the masses $m_i$ and positions ${\bf x}_i$ as well as the matter density perturbations are statistically independent. In this case, it is convenient to use the additive property of the cumulant-generating function of $\bf j$ 
\be
    K({\bf  k}) 
    \equiv \ln \left\langle e^{i {\bf k}\cdot {\bf  j}} \right\rangle = K_{\rm M}({\bf k}) + K_{\rm PBH}({\bf k}) \,,
\ee 
where $K_{\rm M}({\bf  k}) \equiv \ln \left\langle e^{i {\bf k}\cdot {\bf  j}_{\rm M}} \right\rangle$ and $K_{\rm PBH}({\bf  k}) \equiv \ln \left\langle e^{i {\bf k}\cdot {\bf  j}_{\rm PBH}} \right\rangle$ are cumulant generating functions of ${\bf  j}_{\rm M}$, ${\bf  j}_{\rm PBH}$, respectively. The average is taken over all random quantities contained in the expression. The probability distribution of $\bf j$ can then be computed as
\bea\label{jdistr1}
	\frac{\td P }{\td^3 j}
	\equiv \langle \delta({\bf  j} - {\bf  j}_{\rm M} - {\bf j}_{\rm PBH} ) \rangle
	= \int  \frac{\td^3 k}{(2\pi)^3} e^{-i {\bf  k}\cdot{\bf  j} + K({\bf k})} \,.
\eea
Note that the above distribution is two-dimensional since the angular momentum is perpendicular to $\hat{\bf r}$: the cumulant generating function is a function of  ${\bf  k}_{\perp} \equiv {\bf  k}\times \hat {\bf r}$. Statistical isotropy of the surrounding matter further constrains it to be a function of $k_{\perp}$ only.

Let us first consider the matter density fluctuations. The generating function $K_{\rm M}({\bf k})$ is determined by the first two cumulants if the matter density fluctuations are Gaussian, that is
\be
	K_{\rm M}({\bf k}) 
	= -\frac{1}{2}\left\langle ({\bf k}\cdot {\bf j})^2 \right\rangle
	= - \frac{1}{4} \sigma_{j,\rm M}^2 k_{\perp}^2\,, 
\ee
since the mean $\langle{\bf  j}\rangle = 0$ must vanish due to isotropy arguments. The variance $\sigma_{j,\rm M}$ can be estimated by noting that matter density fluctuations induce a tidal field which in Fourier space reads ${\bf T}_{\rm M} = a^{-3} \hat {\bf q} \otimes \hat {\bf q}  \, 4\pi \bar\rho_{\rm M} \delta_{\rm M} (q)$ by Eq.~\eqref{eq:phi}. Using Eq.~\eqref{eq:j} and averaging over orientations then gives~\cite{Raidal:2018bbj,Ali-Haimoud:2017rtz}
\bea\label{eq:sigma_j_M}
    \sigma_{j,\rm M}^2 
&	= \left\langle {\bf  j}_{\rm M}^{2}\right\rangle
    = \left(\frac{0.95 x_0^3}{M}\right)^2 \frac{a^6}{5}\left\langle \tr({\bf T}_{\rm M}\cdot {\bf T}_{\rm M}) - \frac{1}{3}\tr({\bf T}_{\rm M})^2\right\rangle 
    = \frac{6}{5} j_0^2 \frac{\sigma_{\rm M}^2}{f_{\rm PBH}^2} ,
\eea
where $\sigma_{\rm M}^2 \equiv (\Omega_{\rm M}/\Omega_{\rm DM})^2 \left\langle \delta_{\rm M}^2 \right\rangle$ is the rescaled variance of matter density perturbations at the time the binary is formed.\footnote{We will use $\left\langle \delta_{\rm M}^2 \right\rangle = 0.005$ in numerical estimates.} 

The cumulant generating function $K_{\rm PBH}$ accounts for identical and statistically independent contributions from all surrounding PBHs. To send the number of surrounding PBHs to infinity, we will first consider a finite volume and then take the limit $V\to \infty$ while keeping $N/V = n$ fixed:
\footnote{
After plugging in  Eq.~\eqref{j1}, the integral for $K_{\rm PBH}$ takes the form
\bea
    K_{\rm PBH}({\bf  k}) 
    &	= \int \td n(m) \, \int_{|{\bf x}|>y}\frac{\td\Omega}{4\pi}\, \td V(x) \left[\exp\left(\frac{i3mj_0}{\rho_{\rm PBH} V(x)} ({\bf  k}_{\perp} \cdot \hat{\bf x}) \,(\hat {\bf x} \cdot \hat {\bf r}) \right) - 1\right] .
\eea
The spatial integral can be evaluated by making the change of variables $u \equiv z V(y)/V(x)$, with $z \equiv k_{\perp} j_0 m/(\rho_{\rm PBH} V(y))$, and taking the following steps:
\bea \nonumber
&	-z \int\frac{\td^2\Omega}{4\pi}\, \int^{z}_{0} \frac{\td u}{u^2}  \left[\exp\left(i 3 u (\hat{\bf  k}_{\perp} \cdot \hat{\bf x}) \,(\hat {\bf x} \cdot \hat {\bf r}) \right) - 1\right] \\
&	= -z \int^{z}_{0} \frac{\td u}{u^2}  \int^{\pi}_{0} \int^{2\pi}_{0} \frac{\td\cos(\theta) \, \td \phi }{4 \pi} \,   \left[\exp\left(i \frac{3u}{2} \sin^2(\theta)\sin(2\phi) \right) - 1\right] \\
&	= -z \int^{z}_{0} \frac{\td u}{u^2}   \left[\frac{\pi}{2\sqrt{2}} J_{-\frac{1}{4}} \left(\frac{3u}{4}  \right) J_{\frac{1}{4}} \left(\frac{3u}{4}  \right) - 1\right]
	= {}_1F{}_2\left(-\frac{1}{2};\frac{3}{4},\frac{5}{4};-\frac{9z^2}{16}\right) - 1.
\eea
}
\bea
	K_{\rm PBH}({\bf  k}) 
&    = \lim_{\substack{V \to \infty \\ N/V = n}} N \ln \left(\int_{|{\bf x}| > y} \frac{\td^3 x}{V} \frac{\td n(m)}{n} e^{i {\bf k}\cdot {\bf  j}_{1}({\bf x}, m)}\right) \\
&   = \int \td^3 x \, \td n(m) \left(e^{i{\bf k}\cdot {\bf  j}_1({\bf x}, m)} - 1\right) \\
&	= -\bar{N}(y)\int \frac{\td n(m)}{n} \,F\left( \frac{m}{\langle m \rangle} \frac{1}{\bar{N}(y)} j_0 k_{\perp} \right) \,,	
\eea
where $F(z) = {}_1F{}_2\left(-1/2;3/4,5/4;-9z^2/16 \right) - 1$ and ${}_1F{}_2$ is the generalised hypergeometric function. It asymptotes to $F(z) \sim z - 1$ when $z \to \infty$ and $F(z) \sim 3z^2/10$ when $z \to 0$. These limits correspond to $\bar{N}(y) \ll 1$ and $\bar{N}(y) \gg 1$, respectively. In particular, the $\bar{N}(y) \gg 1$ case recovers the Gaussian limit $K_{\rm PBH}({\bf k}) \sim - \frac{1}{4} \sigma_{j,\rm PBH}^2 k_{\perp}^2$, with the variance given by
\be\label{eq:sigma_j_PBH}
    \sigma_{j,\rm PBH}^2 
    = \frac{6}{5} j_0^2 \frac{\langle m^2 \rangle}{\langle m \rangle^2} \frac{1}{\bar{N}(y)} .
\ee
The variance diverges when $\bar{N}(y) \to 0$. For this reason, it is not possible to use the central limit theorem to compute the angular momentum distribution even though it arises from an infinite sum of independent identical random variables. We also remark that a simple parallel can be drawn between the variances~\eqref{eq:sigma_j_M} and~\eqref{eq:sigma_j_PBH} by noting that the variance of PBH density fluctuations in a volume $V(y)$ is $\langle m^2 \rangle/(\langle m \rangle^2 \bar{N}(y))$ and has thus an analogous interpretation to the factor $\sigma_{\rm M}^2/f_{\rm PBH}$ in Eq.~\eqref{eq:sigma_j_M}.

\begin{figure}[t]
\begin{center}
\includegraphics[width=0.75\textwidth]{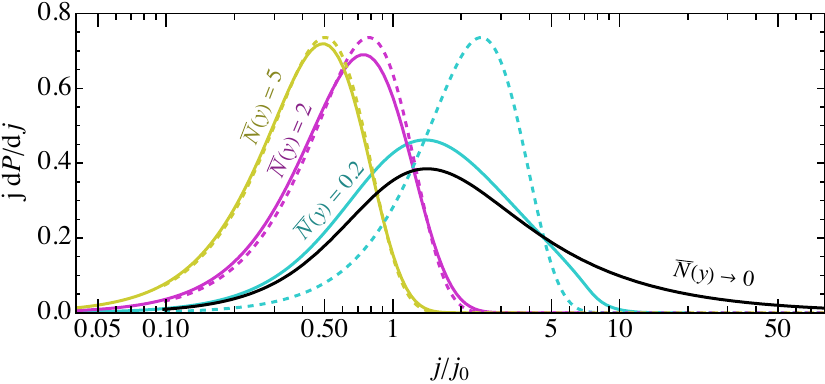}
\caption{Angular momentum distributions of PBH binaries after formation \eqref{eq:Pj} for different $\bar{N}(y)$ are shown by solid lines. The dashed lines show the corresponding Gaussian approximation~\eqref{eq:Pj_limits}. A monochromatic mass function is assumed in the figure.}
\label{fig:Pj}
\end{center}
\end{figure}

Due to isotropy, it is sufficient to consider the distribution of $j$ instead of $\bf j$. The cumulant generating functions imply the angular momentum distribution~\eqref{jdistr1} is given by
\bea \label{eq:Pj}
    j \frac{\td P}{\td j}  
    = \int^{\infty}_{0}  \td u \,u  J_{0}(u ) \exp\bigg[ &-\bar{N}(y) \int \frac{\td m}{m} \psi(m)\,F\left(u \frac{m}{\langle m \rangle} \frac{1}{\bar{N}(y)} \frac{j_0}{j}\right) \\ 
    &- u^{2} \frac{3}{10}\frac{\sigma_{\rm M}^{2}}{f_{\rm PBH}^{2}}\frac{j_{0}^{2}}{j^{2}}\bigg]\,.
\eea

Depending on the size of the empty region surrounding the PBH pair, this distribution interpolates between two asymptotic cases
\be \label{eq:Pj_limits}
    j \frac{\td P}{\td j}
    = \left\{
    \begin{array}{ll}
        \dfrac{ j^2/j_0^2}{(1+ j^2/j_0^2)^{3/2}},     & \quad \bar{N}(y) \ll 1 \quad \mbox{(other PBHs arbitrarily close)},\\
        \dfrac{2j^2}{\sigma_j^2} e^{-j^2/\sigma_j^2}, & \quad \bar{N}(y) \gg 1 \quad \mbox{(Gaussian limit)},
    \end{array}
    \right.
\ee
where 
\be\label{eq:sigma_j}
    \sigma_j^2 
    \equiv \sigma_{j,\rm M}^2+ \sigma_{j,\rm PBH}^2 
    = \frac{6}{5} j_0^2 \left( \frac{1}{\bar{N}(y)} \frac{\langle m^2 \rangle}{\langle m \rangle^2}  + \frac{\sigma_{\rm M}^2}{f_{\rm PBH}^2} \right)
\ee
is the total variance of the distribution. The angular momentum distribution \eqref{eq:Pj} together with the Gaussian approximation is shown in Fig.~\ref{fig:Pj} for a monochromatic mass function. It can be seen that the Gaussian approximation works well already for $\bar{N}(y) = 2$, while the $\bar{N}(y) \to 0$ limit becomes a good approximation when $\bar{N}(y) \lesssim 0.2$.

The appearance of $\psi(m)$ in \eqref{eq:Pj} implies a non-trivial dependence on the mass function. However, the dependence on $\psi(m)$ vanishes completely \eqref{eq:Pj_limits} in the $\bar{N}(y) \ll 1$ limit and can be reduced to the relative width of the mass distribution in the Gaussian $\bar{N}(y) \gg 1$ case. 

\begin{figure}[t]
\begin{center}
\includegraphics[width=0.98\textwidth]{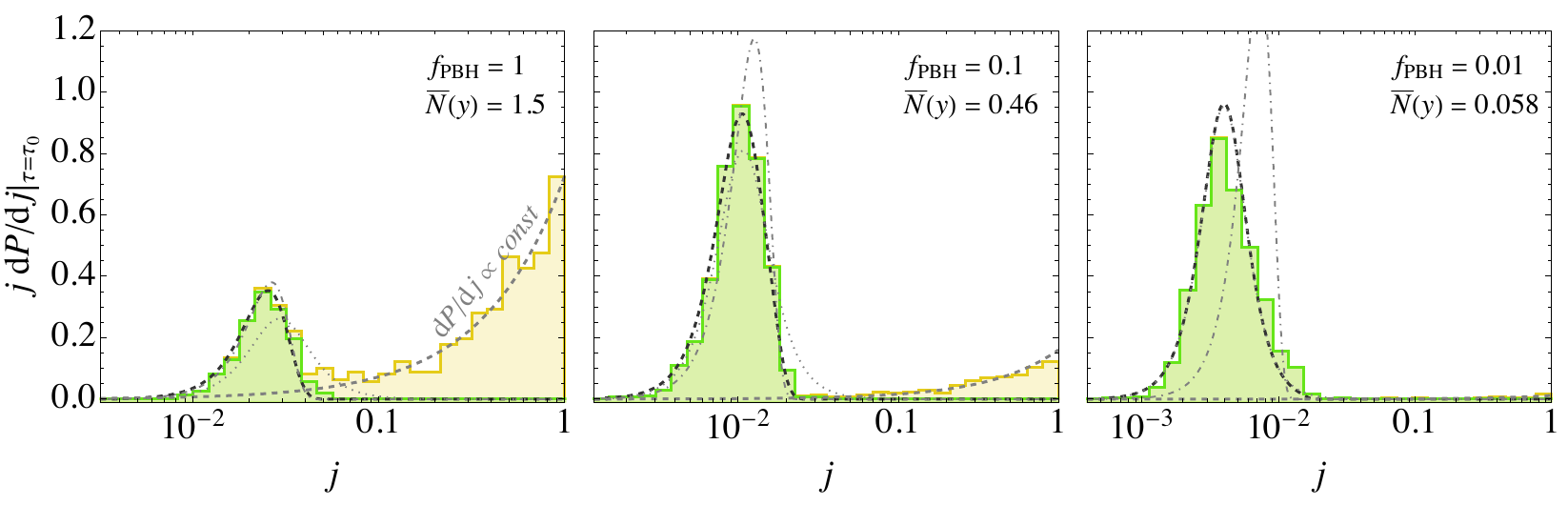}
\caption{Distribution of angular momenta at $a  = 3 \, a_{\mathrm{eq}}$ of PBH binaries for different $f_{\rm PBH}$ and a monochromatic mass function with $m_c = 30 \Msun$. The initial configurations were chosen in a way that the PBH pair would form a binary that merges today if unperturbed. The perturbed and unperturbed binaries are coloured in yellow and green respectively. The thick dashed line shows the distribution~\eqref{eq:Pj_tau} with $\bar{N}(y)$ estimated from Eq.~\eqref{eq:nVy_collapse}. The thin dashed curves show limiting cases. The thick grey dashed curve compares the distribution of disrupted binaries to $\td P/\td j \propto 1$. Based on simulation data from Ref.~\cite{Raidal:2018bbj}.}
\label{fig:dist_j}
\end{center}
\end{figure}

Eq.~\eqref{eq:Pj} is in excellent agreement with numerical simulations as illustrated by  Fig.~\ref{fig:dist_j}. It shows the angular momentum distributions obtained from $N$-body simulations of PBH binaries expected to merge within the age of the Universe $\tH$ (green region) and compares them to the distribution predicted by Eq.~\eqref{eq:Pj} (black dashed curve). Note that the distribution~\eqref{eq:Pj} gives a conditional probability assuming a fixed initial separation $x_0$. The corresponding distribution conditioned on a fixed coalescence time $t$ is
\bea\label{eq:Pj_tau}
    \frac{\td P(j|\tau = t)}{\td j} 
    \propto \left.   j_0(x_0) \frac{\td P(j|x_0)}{\td j}   \right|_{\tau(r_{a}(x_0),j)=t} \,,
\eea
where we dropped prefactors independent of $j$. Analogously to the distribution $j\td P(j|x_0)/\td j$ that depends on $j$ only through the combination $j/j_0$, it is possible to define $j_\tau$ so that $\td P(j|\tau)/\td j$ would depend on $j$ through the combination $j/j_\tau$ only, where $j_\tau$ gives an order of magnitude estimate of the dimensionless angular momentum of initial binaries merging today. Since $j/j_0|_{\tau=t} =  (j/j_\tau)^{37/16}$, we have that
\be\label{eq:j_tau}
    j_\tau 
    =  1.7 \times 10^{-2} \left[\frac{t}{\tH}\right]^{\frac{3}{37}} \left[\frac{M}{ \Msun}\right]^{\frac{5}{37}} (4\eta)^{\frac{3}{37}} f_{\rm PBH}^{\frac{16}{37} }\,.
\ee
The dependence on both the mass and the coalescence time is very mild, so it is safe to assume that PBHs around the stellar mass range had initial eccentricities around $\mathcal{O}(10^{-2})$ or lower when $f_{\rm PBH} \ll 1$.

\subsubsection{Merger rate and binary disruption by nearest PBHs}
\label{sec:RE2}

The binary may be disrupted shortly after formation by becoming bound to a nearby PBH or later by interacting with PBHs in compact DM haloes, which are more likely to form in PBH cosmologies. The first possibility is accounted by an appropriate choice of $\bar{N}(y)$ and it is typically the dominant effect when $f_{\rm PBH} \ll 1$. When $f_{\rm PBH} \approx 1$, however, the binary is likely to be perturbed by later encounters due to the enhanced small-scale structure of PBH DM. This effect will be addressed separately in Sec.~\ref{sec:disruption_haloes}.

Since $j_0 = \mathcal{O}(10^{-2})$ for binaries merging today, the early PBH binaries tend to be extremely eccentric at formation. Since the coalescence time scales as $\tau \propto j^7$, even a $\mathcal{O}(1)$ increase in $j$ will increase the coalescence time by several orders of magnitude. Therefore, even mild interactions can effectively remove the binary from the population of binaries merging today. It is certainly possible that some of the disrupted binaries may contribute to the present merger rate, but they must evolve from initially more compact PBH systems to merge within the age of the Universe after being perturbed. This scenario constitutes an entirely different formation channel -- the early three-body channel -- discussed in Sec.~\ref{sec:RE3}.

\begin{figure}[t]
\begin{center}
\includegraphics[width=0.98\textwidth]{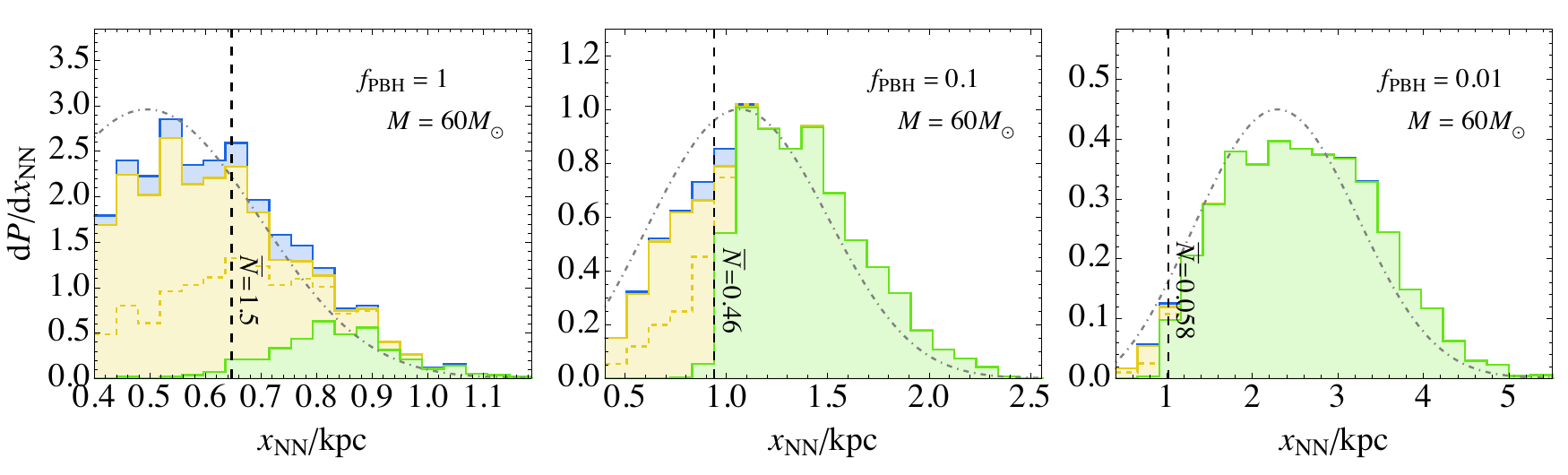}
\caption{Binary disruption at $a  = 3 \, a_{\rm eq}$ depending on the initial comoving distance $x_{\rm NN}$ of the PBH nearest to the binary for $30 \Msun$ PBHs and different $f_{\rm PBH}$. The blue region indicates simulations in which the initial pair did not form a binary. The yellow region shows pairs that evolved into a binary that was later perturbed but remained bound. Binaries above the yellow dashed line swapped at least one of their components. The green region shows undisrupted binaries. The vertical dashed line indicates the estimate Eq.~\eqref{eq:nVy_collapse}. The numerical simulations evolve a PBH pair estimated to evolve into a binary that would merge at the present time if not disrupted. Based on simulation data from Ref.~\cite{Raidal:2018bbj}.
}
\label{fig:dist_nn}
\end{center}
\end{figure}

To determine $\bar{N}(y)$, consider collisions between the binary and the PBH initially closest to it. Let $x_{\rm NN}$ denote the initial comoving distance of the nearest neighbour. Since $y$ sets a lower bound on viable values of $x_{\rm NN}$ for binary formation, we must find the minimal $x_{\rm NN}$ which does not lead to the disruption of the binary after the nearest neighbour decouples from the Hubble flow. This estimate can be made following the reasoning of Sec.~\ref{sec:decoupling}. The region surrounding the binary corresponds to an effective matter density fluctuation $\delta_{\rm NN} \approx (M-\rho_{\rm PBH} V(x_{\rm NN}))/(\rho_{\rm M} V(x_{\rm NN}))$. Given that such configurations are expected to collapse at $a \approx a_{\rm eq}/\delta_{\rm NN}$ we estimate that, at a given redshift, the binary has collided with its neighbour if $x_{\rm NN} < y$, when
\be\label{eq:nVy_collapse}
	\bar{N}(y) \approx \frac{M}{\langle m\rangle} \frac{f_{\rm PBH}}{f_{\rm PBH} + a_{\rm eq}/a} \,.
\ee
The condition~\eqref{eq:nVy_collapse} agrees well with numerical $N$-body simulations of PBH binary formation in the early Universe~\cite{Raidal:2018bbj}. This is illustrated in Fig.~\ref{fig:dist_nn}, which shows that \eqref{eq:nVy_collapse} predicts the interface between perturbed (yellow region) and unperturbed (green region) binaries very well when $f_{\rm PBH} \ll 1$. When $f_{\rm PBH} = 1$, about half of the binaries with $x_{\rm NN} > y$ are seen disrupted. As we will show in Sec.~\ref{sec:disruption_haloes}, the early formation of PBH clusters in such PBH cosmologies~\cite{Inman:2019wvr} can account for this effect. The estimate Eq.~\eqref{eq:nVy_collapse} was also used when comparing numerically obtained angular momentum distributions to our analytic predictions in Fig.~\ref{fig:dist_j}. In that case, an excellent agreement can be observed also when $f_{\rm PBH} = 1$. 

When $a \gg a_{\rm eq}$ and for narrow mass functions, we have that $\bar{N}(y) \approx 2$ implying that the PBH pair should be positioned in an underdense region. This condition is, however, too strong when the PBH abundance is small $f_{\rm PBH} \ll 1$. In this case, it is sufficient to demand that the binary survives until the non-linear structure formation of CDM haloes sets in. Since this happens when $a/a_{\rm eq} = 1/\sigma_M$, we have that
\be\label{eq:nVy}
	\bar{N}(y) \approx \frac{M}{\langle m \rangle}\frac{f_{\rm PBH}}{f_{\rm PBH} + \sigma_{\rm M}}.
\ee
The binary is expected to evolve without much interference after that point since the disruption of PBH binaries in the standard late CDM haloes is very unlikely~\cite{Ali-Haimoud:2017rtz}. This point is illustrated further in section~\ref{sec:disruption_haloes}, where we show that, even when considering the enhanced small-scale structures in PBH cosmologies, PBH encounters are expected to take place in relatively small haloes containing less than $\mathcal{O}(10^4)$ PBHs which are formed at high redshifts.

A potential shortcoming of the condition~\eqref{eq:nVy} is that it accounts for the masses of the surrounding PBHs on average. Thus it can fail when the mass distribution of PBHs spans several orders of magnitude. For example, it can underestimate the suppression when the nearest PBH is much heavier than the total mass of the binary since heavier PBHs are more likely to disrupt light binaries even if their distance is large. On the other hand, it can overestimate the disruption of heavy binaries by light PBHs surrounding it. In extreme cases, the surrounding binaries can be too light to significantly alter the orbit of the binary. In that case, excluding initial configurations where a light PBH is close to binary is unjustified.

Having determined the characteristics of the first binaries and constraints to their environment, we can proceed to evaluate the merger rate density \eqref{dR_early} $\td R_{E2} = S_{\rm E}S_{\rm L} \td R^{(0)}_{E2}$. The unsuppressed contribution $\td R^{(0)}_{E2}$ is evaluated by integrating Eq.~\eqref{dR_early_y} in the $\sigma_M \to 0$ limit and using $y\to 0$ asymptotic~\eqref{eq:Pj} of the angular momentum distribution, 
\bea
    \frac{\td R^{(0)}_{E2}(t)}{\td m_1 \td m_2}
&	=  \frac{1.4}{t} \left( \frac{t\eta  M^{14}}{f_{\rm PBH}^{7} \rho_{M}^{11}}\right)^{\frac{3}{37}} \td n(m_{1})\td n(m_{2}) \\
&	\approx \frac{1.6 \times 10^{6}}{\Gpc^{3} \yr} \, f_{\rm PBH}^{\frac{53}{37}} \eta^{-\frac{34}{37}}  \left(\frac{M}{\Msun}\right)^{-\frac{32}{37}} \left(\frac{t}{\tH}\right)^{-\frac{34}{37}}  \, \frac{\psi(m_1) \psi(m_2)}{m_1 m_2} .
\eea
The suppression factor $S_{\rm E}$ gives the remaining contribution\footnote{Integrating \eqref{dR_early_y} over $x_0$ and $u$ in Eq.~\eqref{eq:Pj} can be simplified by changing $x_0$ to $v = u j_0/j(\tau,x_0)$, which helps to factor out parts of the integral.}
\bea\label{def:S}
    S_{\rm E}
    = \frac{e^{-\bar{N}(y)}}{\Gamma(21/37)} \int \td v \, v^{-\frac{16}{37}} \exp\bigg[ 
&   -\bar{N}(y) \langle m \rangle \int \frac{\td m}{m} \psi(m) F\left(\frac{m}{\langle m \rangle} \frac{v}{\bar{N}(y)}\right)  \\
&   - \frac{3\sigma_{\rm M}^{2} v^{2}}{10 f_{\rm PBH}^{2}}  
    \bigg] \, ,
\eea
with $\bar{N}(y)$ estimated from \eqref{eq:nVy}. 

The suppression factor includes the effect of matter perturbations. In particular, when $f_{\rm PBH} \ll \sigma_{M}$, the angular momentum of the early binaries is generated mainly by matter perturbations (compare Eqs.~\eqref{eq:sigma_j_M} and \eqref{eq:sigma_j_PBH}) instead of the surrounding PBHs. As a result, the merger rate scales $f_{\rm PBH}^2$ instead of $f_{\rm PBH}^{53/27}$. This slightly stronger dependence causes the merger rate to drop faster when $f_{\rm PBH}$ is decreased. Since in this regime $\td R^{(0)}_{E2}(t) \propto \sigma_{\rm M}^{-21/74}$ a larger $\sigma_{\rm M}$, that is, more pronounced matter perturbations, will lead to a smaller merger rate.

In practice \eqref{def:S} must be evaluated numerically, which can be quite computationally expensive. A good analytic approximation can be constructed by considering the $\bar{N}(y) \to 0$ and $\bar{N}(y) \to \infty$ asymptotic of the angular momentum and noting that the following inequalities hold~\cite{Raidal:2018bbj}
\be
    S_{\rm E,\rm min} \leq S_{\rm E} \leq S_{\rm E,\rm max} < 1 \,,
\ee
where the $\bar{N}(y) \to 0$ and $\bar{N}(y) \to \infty$ limits are
\bea
    S_{\rm E,\rm max}
&    = \left(\frac{5f_{\rm PBH}^{2}}{6 \sigma_{\rm M}^{2}}\right)^{\frac{21}{74}} U\left(\frac{21}{74},\frac{1}{2},\frac{5f_{\rm PBH}^{2}}{6 \sigma_{\rm M}^{2}}\right) 
\,,\\
    S_{\rm E,\rm min} 
&	= \frac{\sqrt{\pi}(5/6)^{21/74}}{\Gamma(29/37)}  \left[\frac{\langle m^2\rangle/\langle m\rangle^2}{\bar{N}(y)} + \frac{\sigma_{\rm M}^2}{f_{\rm PBH}^2} \right]^{-\frac{21}{74}} e^{-\bar{N}(y)} \,,
\eea
where $U$ denotes the confluent hypergeometric function. The factor $S_{\rm E}$ can be approximated as
\be\label{eq:S1_approx}
    S_{\rm E} \approx \frac{\sqrt{\pi}(5/6)^{21/74}}{\Gamma(29/37)}
    \left[\frac{\langle m^2\rangle/\langle m\rangle^2}{\bar{N}(y) + C } + \frac{\sigma_{\rm M}^2}{f_{\rm PBH}^2}\right]^{-\frac{21}{74}} e^{-\bar{N}(y)} \,,
\ee
where the factor $C$ is fixed by demanding that $S_{\rm E} \to S_{\rm E,\rm max}$ in the limit $\bar{N}(y) \to 0$,
\be
    C 
    =  f_{\rm PBH}^{2} \frac{\langle m^2\rangle/\langle m\rangle^2}{\sigma_{\rm M}^{2}} \left\{ \left[ \frac{\Gamma(29/37)} {\sqrt{\pi}} U\left(\frac{21}{74},\frac{1}{2},\frac{5f_{\rm PBH}^{2}}{6 \sigma_{\rm M}^{2}}\right) \right]^{-\frac{74}{21}} - 1 \right\}^{-1} \,.
\ee
This approximation is accurate within 7\% for log-normal mass function with widths $\sigma \leq 2$. Another approximation, that is accurate within 2\% but retains one numerical integral, has been constructed in Ref.~\cite{Hall:2020daa}.

\subsection{Disruption of early binaries in DM haloes}
\label{sec:disruption_haloes}

After its formation, the binary may get perturbed by encounters with a third PBH. Such encounters are likely in very dense haloes. We can estimate the probability of this by computing the rate of encounters that significantly alter the angular momentum of the binary. However, the small haloes are also unstable and experience gravothermal instabilities within the age of the Universe. This can lead to core collapse and the eventual evaporation of the PBH cluster. In the former case, the density can increase significantly and all binaries in such haloes are likely to get perturbed. In the following, we first estimate the rate of encounters that perturb the binaries and then the probability that the binary is within a halo that goes through gravothermal collapse before the binary merges.

Consider an early binary whose coalescence time initially equals the age of the Universe, $\tau = \tH$ and a third PBH of mass $m_3$ approaching the binary. In a hyperbolic encounter,\footnote{See Sec.~\ref{sec:late} for details of the hyperbolic encounters.} the velocity of the third body of mass $m_3$ at the distance of closest approach $r_p$ is $v_p = b v_{\rm rel}/r_p$, where $v_{\rm rel}$ denotes the relative velocity of the third body and $b$ the impact parameter that is related to $r_p$ by $b^2 = r_c^2 + 2(M+m_3) r_c/v_{\rm rel}^2$. We can estimate the timescale of the interaction between the binary and the third body as $t_p \sim r_p/v_p$. The tidal torque caused by the third body is given by ${\bf T}_p =  m_3 ({\bf 1} - 3\hat {\bf r}_p \otimes \hat {\bf r}_p )/r_p^3$ and the resulting change in the binary angular momentum can be estimated as $\Delta {\bf L} \simeq t_p {\bf r}_a\times ({\bf T}_p \cdot {\bf r}_a)$. This gives
\be
    \Delta j \simeq \frac{m_3 r_a^{3/2}}{\sqrt{M} b v_{\rm rel} r_p} \,.
\ee
The binary gets perturbed if $\Delta j$ is comparable to the initial angular momentum. For the early binaries, we can estimate the initial angular momentum by $j_\tau$ given in Eq.~\eqref{eq:j_tau}. So, encounters with the impact parameter 
\be
    b \lesssim \frac{[m_3(M+m_3)]^{1/3} r_a^{1/2}}{j_\tau^{1/3} M^{1/6} v_{\rm rel}} \equiv b_{\Delta j > j_{\tau}}
\ee
would perturb the binary. The corresponding interaction cross section is $\sigma_{\Delta j > j_{\tau}} = \pi b_{\Delta j > j_{\tau}}^2$.

To estimate how large the rate of such processes is, let us consider early PBH haloes of $N$ PBHs. Such haloes form when $a_c \equiv (1+z_c)^{-1} \approx a_{\rm eq} \sqrt{N}/f_{\rm PBH}$, where $a_{\rm eq}$ denotes the matter-radiation equality. If such haloes are virialized, the velocity dispersion is $\sigma_v^2 \approx M_{\rm H}/R$ where $R$ denotes the virial radius and $M_{\rm H} = \langle m\rangle N/f_{\rm PBH}$ the mass of the halo.\footnote{Here we assume that the fraction of DM in PBHs in haloes matches $f_{\rm PBH}$. Ref.~\cite{Inman:2019wvr} showed that, especially in the early Universe, this fraction could be larger.} The average matter density in these haloes is, following the Press-Schechter theory, given by $\rho = 3M_{\rm H}/(4\pi R^3) \approx 18 \pi^2 \rho_{c} a_c^{-3}$, where $\rho_c$ is the critical comoving density. So, the timescale of processes that perturb the binary in such haloes can be estimated as
\be
    t_p \simeq \left[\frac{\rho}{\langle m \rangle} \langle v_{\rm rel} \sigma_{\Delta j > j_{\tau}} \rangle \right]^{-1} \simeq 1 \,{\rm Gyr} \,\left[\frac{N}{1000}\right]^{\frac{19}{12}} f_{\rm PBH}^{-\frac{397}{222}} \left[\frac{\langle m\rangle}{\Msun}\right]^{-\frac{10}{111}},
\ee
where used Eq.~\eqref{eq:tau} to translate $r_a$ into $\tau$, took $\tau = \tH$, replaced all masses by $\langle m \rangle$ and the relative velocity by the velocity dispersion $\sigma_v$. With this rough estimate, we find that for $\langle m \rangle = \mathcal{O}(10M_\odot)$ and $f_{\rm PBH}=1$ binaries in haloes with $N\lesssim 5800$ get perturbed before today. Moreover, by definition, the haloes need at least $N=3$, implying that binaries are likely not perturbed when $f_{\rm PBH} \lesssim 10^{-3}$.

Next, let us estimate the effect of gravothermal instabilities of small haloes. We estimate this using the characteristic time scale of core collapse, $t_{\rm cc} \geq 18 t_{\rm r}$, where $t_{\rm r}$ denotes the relaxation time~\cite{Quinlan:1996bw}:
\be
	t_{\rm r} = 0.065 \frac{\sigma_v^3}{m \rho \ln \Lambda} 
	\approx 2 \kyr \frac{N^{7/4}}{f_{\rm PBH}^{5/2} \ln \Lambda} \,.
\ee
By approximating the Coulomb logarithm as $\ln\Lambda \approx \ln (N/f_{\rm PBH})$ and requiring $18t_{\rm r} < \tH$ we get that haloes with
\be\label{eq:Nc}
	N \leq 1500 f_{\rm PBH}^{10/7} \ln\Lambda^{4/7} \equiv N_c
\ee
experience gravothermal collapse before today (see also~\cite{Afshordi:2003zb}). This indicates that, if $f_{\rm PBH}=1$, then binaries are perturbed in haloes with $N < 5300$. All binaries survive if $f_{\rm PBH} \lesssim 0.005$. These numbers are very similar to those found above by estimating the rate of the encounters that would perturb the binaries assuming that the haloes survive.

Let us briefly consider the hardness of the early binaries. In the largest haloes in which binaries are likely disrupted according to Eq.~\eqref{eq:Nc}, the velocity dispersion is
\be
    \sigma_v 
    \approx 0.8 \frac{\rm km}{\rm s}\, f_{\rm PBH}^{\frac16} \left[\frac{\langle m\rangle}{\Msun}\right]^{\frac{1}{3}} N^{\frac{1}{12}} 
    \approx 1.8 \frac{\rm km}{\rm s}\, f_{\rm PBH}^{\frac27} \left[\frac{\langle m\rangle}{\Msun}\right]^{\frac{1}{3}}  \, .
\ee
On the other hand, the characteristic binding energy of binaries merging within the age of the Universe is
\be
    \frac{E_{\rm bin}}{\mu} \approx \left( 3.4 \, \frac{\rm km}{\rm s} \,\eta^{-\frac18} \left[\frac{j}{0.01}\right]^{\frac78}  \left[\frac{M}{\Msun}\right]^{\frac{1}{8}}  \right)^2\,.
\ee
Therefore, the binaries can be considered hard as encounters with other PBHs are unlikely to ionise them within environments where encounters are frequent enough.

\begin{figure}[t]
\centering
\includegraphics[width=0.6\textwidth]{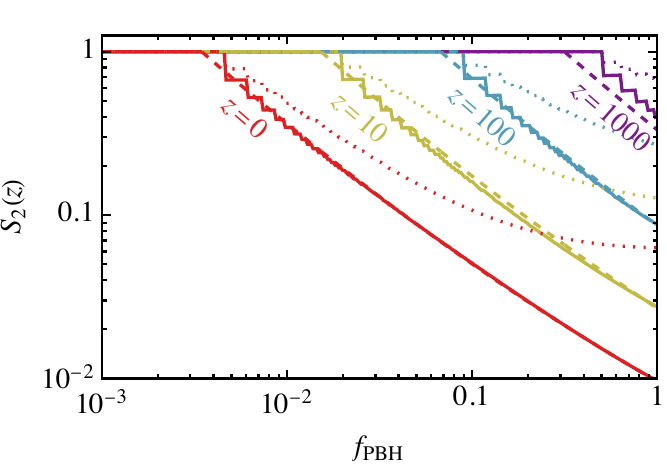}
\caption{The suppression factor of merger rate for the early two-body channel as a function of PBH abundance at different redshifts. The dashed curves show the approximation~\eqref{eq:S2a} and the dotted curves show the suppression factor neglecting the subhalos.}
\label{fig:S2}
\end{figure}

Assuming that all of the binaries in haloes of size $N \leq N_c(z)$ get perturbed, we can estimate the suppression of the merger rate by estimating the probability that the binary is in such halo. Including both isolated haloes and subhaloes, the suppression factor, shown in Fig.~\ref{fig:S2}, is given by~\cite{Vaskonen:2019jpv} 
\be
    S_{\rm L}(z) = 1 - \sum_{N=3}^{N_{c}(z)} \left[p_{N}(z_c) + \sum_{N'>N_{c}(z)} p_{N,N'}(z_c) \right] \,,
\ee
where $z_c = f_{\rm PBH} z_{\rm eq}/\sqrt{N_c}$ denotes the redshift at which the haloes of $N_c$ PBHs form and $z_{\rm eq}$ denotes the redshift of matter-radiation equality. The probability of the binary being part of a halo containing $N$ PBHs at redshift $z$ is given by
\be
    p_{N}(z) = \frac{N^{-1/2}e^{-N/N^{*}(z)}}{\sum_{N>2} N^{-1/2}e^{-N/N^{*}(z)}} \,,
\ee
and the probability of the binary being part of a subhalo containing $N$ PBHs inside a halo of $N'$ by
\bea
    p_{N,N'}(z) &= p_{N'}(z) \,\frac{N^{-1/2}e^{-N/N^{*}(z)}}{\sum_{N = 2}^{N'} N^{-1/2}e^{-N/N^{*}(z)}} \,.
\eea
Here $N^*(z)$ is the characteristic number of PBHs in a halo at redshift $z$, and can be estimated by~\cite{Inman:2019wvr}
\be
    N^*(z) = \left[\ln(1+\delta^*) - \delta^*/(1+\delta^*)\right]^{-1} 
\ee
with
\be
    \delta^* \approx 
    1.69 \left[f_{\rm PBH}\, {}_1F{}_2\left((1-\sqrt{21})/4,(1+\sqrt{21})/4;1;-(z_{\rm eq}+1)/(z+1) \right) \right]^{-1} \,.
\ee

The numerically evaluated suppression factor $S_{\rm L}(z)$ can be approximated by~\cite{Vaskonen:2019jpv}
\be \label{eq:S2a}
    S_{\rm L}(t) \approx 
    \min\left\{1,\,0.01 \left[ \left(\frac{t}{\tH}\right)^{0.44} f_{\rm PBH} \right]^{-0.65} e^{0.03 \ln^2 \left[\left(\frac{t}{\tH}\right)^{0.44} f_{\rm PBH}\right]}\right\}\,,
\ee
which, as seen in Fig.~\ref{fig:S2}, fits the numerical estimate well. The plot also shows that for large $f_{\rm PBH}$ (e.g. $f_{\rm PBH} > 0.03$ at $z=0$) the dominant contribution to the suppression arises from binary disruption in subhalos. We remark that this suppression factor might be affected by phenomena, such as tidal stripping, which can modify subhalo evolution. So far, DM subhaloes in PBH cosmologies have not been studied in detail and thus this term might be subject to unknown systematic uncertainties.

\subsection{The early three-body channel: merger rate of perturbed binaries}
\label{sec:RE3}

As discussed above, a sizeable fraction of PBHs will form binaries after they decouple from the Hubble flow. However, many of them can be perturbed by interactions with other PBHs. The population of such binaries will not share the characteristics of the unperturbed binaries evolving in isolation. This was sharply illustrated by the simulated angular momentum distribution shown in Fig.~\ref{fig:dist_j}, which revealed a peaked distribution for unperturbed binaries, but a nearly flat one ($\td P/\td j \approx 1$) for perturbed ones. The perturbed binaries tend to be hard so the interactions with other PBHs are likely not break them apart. Since they are expected to be still around, many of them could contribute to the present merger rate.

Most perturbed binaries in the early Universe that would merge today would have formed from PBH pairs that initially evolved into an eccentric binary with a very short coalescence time. Let us first make a few order-of-magnitude estimates to understand the formation mechanism. Consider a compact PBH pair that decouples from the Hubble flow. Its expected initial angular momentum due to tidal forces is $j = \mathcal{O}(10^{-2})$. After being perturbed we expect that $j = \mathcal{O}(1)$, that is, the angular momentum grows by about two orders of magnitude. Assuming that this process formed a binary that merges today and given that $\tau \propto j^{7}$, the initial coalescence time would have been 14 orders of magnitude shorter than the age of the Universe, that is, about an hour. Such binaries would not have had time to interact with other PBHs before the merger. 

It follows that perturbed binaries contributing to the present PBH merger rate are formed from PBH pairs residing in a tidal field that induces angular momenta $j \gg \mathcal{O}(10^{-2})$, which tend to be larger than for unperturbed binaries discussed in the last section. Such a tidal field implies the presence of a third PBH at a distance closer than the average separation. The presence of a nearby 3rd PBH means that we are looking at a 3-body system that eventually evolves into a binary after the third body is ejected. The binding energy of the binary is comparable to the binding energy of the 3-body system and its angular momentum can be estimated to follow the distribution
\be\label{eq:Pj_pert}
    \frac{\td P(j)}{\td j} = \gamma j^{\gamma-1} ,
\ee
with $\gamma \in [1,2]$. The limiting case $\gamma = 1$ is seen in numerical simulations of early Universe binary formation~\cite{Raidal:2018bbj} and is shown in Fig.~\ref{fig:dist_j} by the yellow histogram. The other limit, $\gamma = 2$, corresponds to the thermal distribution~\cite{1919MNRAS..79..408J}. However, we note that gravitating many-body systems are unstable and will not thermalize completely, so it is not clear if a thermal $\gamma=2$ distribution can be realized. Also, assuming that the angular momentum distribution $P(j)$ is independent of the initial angular momentum agrees with numerical studies of binary-single body collisions~\cite{1993ApJ...415..631S,Fregeau:2004if}. 

With these considerations in mind, we can estimate the merger rate of perturbed binaries. Consider three PBHs with masses $m_i$, $i\in \{1,2,3\}$ that are of the similar. We assume that the Hubble decoupling of the 3-body system is hierarchical\footnote{This assumption is likely to hold when all comoving separations are below the average, in which case hierarchical 3-body configurations would be more likely than configurations in which all distances are comparable.} so that the PBHs "1" and "2" at a comoving separation $x$ will decouple first at 
\be
    a(t_{12}) \approx 
    a_{\rm eq} \frac{N(x)}{f_{\rm PBH}} \frac{\langle m \rangle}{m_1+m_2} ,
\ee  
and the third PBH at a comoving distance $y$ from the centre of mass of the pair will decouple at
\be
    a(t_p) \approx a_{\rm eq} \frac{N(y)}{f_{\rm PBH}} \frac{\langle m \rangle}{m_1+m_2+m_3}\, .
\ee  
During radiation domination
\be \label{eq:t_p}
    t_{\rm p} 
    = 65.3 \, \kyr \times
    \left(\frac{\langle m\rangle}{m_1+m_2+m_3}\right)^2 \frac{N(y)^2}{f_{\rm PBH}^2}  \,.
\ee
As discussed above, the coalescence time $\tau_{\rm i}$ of the binary "1+2" can be short. In the following, we will consider systems with
\be~\label{eq:taubound}
    \tau_{\rm i} > 20 t_{\rm p}.
\ee
The binding energy of eccentric binaries scales as $E(t) = E_0 (1-t/\tau)^{-2}$~\cite{Peters:1964zz}, so this condition is equivalent to demanding that the binding energy is affected by less than 11\% by GW emission. We also note that, during matter-domination, Eq.~\eqref{eq:t_p} overestimates $t_{\rm p}$, so~\eqref{eq:taubound} would lead to a lower merger rate estimate. 

With hierarchical collapse, the coalescence time of the initial "1+2" binary depends on its angular momentum $j_{12}$ which is set mostly by the tidal torque from the third PBH. By Eqs. \eqref{j1} and \eqref{def:j0}, it is given by 
\be
    j_{12} = 1.4 \frac{m_3}{m_1+m_2} 
    \frac{\bar N(x)}{\bar N(y)} |\sin(2\theta)|,
\ee 
where $\theta$ denotes the angle between the line joining the "1+2" pair and the line joining the center of mass of the pair and the third PBH. The order of magnitude of the angular momentum generated by the remaining PBHs is of the order of the width of the distribution~\eqref{eq:sigma_j}, that is, $\sigma_{j} \approx 0.5 \bar N(x) /\sqrt{\bar N(y)}$. If $\bar N(y) \ll 1$, then $j_{12} \gg \sigma_{j}$ and we can neglect the contribution from the remaining PBHs.

In a hierarchical setup, the binding energy of the 3-body system is determined mostly by the binding energy of the initial binary. The comoving density of initial "1+2" binaries with per binding energy interval $(E,E+\td E)$ is 
\be \label{eq:niE0}
    \frac{\td n_i(E,m_i)}{\td E} = \int \td n_{\rm 3} \, P(E,\tau_i \geq 20 t_p|x,y,\theta,m_i) \,,  
\ee
where
\be\label{eq:n_b3}
    \td n_{\rm 3} = \frac{n}{2} \frac{\td n(m_1)}{n} \frac{\td n(m_2)}{n} \frac{\td n(m_3)}{n} \td \bar N(x)  \td \bar N(y) e^{-\bar N(y)} \frac{\td \cos \theta}{2}
\ee
gives the density of 3-body configurations yielding the initial binary. It generalizes Eq.~\eqref{eq:n_b}. The probability that the "1+2" pair produces a binary with binding energy $E$ that satisfies the bound~\eqref{eq:taubound} is
\bea \label{eq:Pconf}
 	P(&E,\tau_i \geq 20 t_p|x,y,\theta,m_i) \\
	&=\delta(E - E(x))\Theta(\tau_{\rm i}(E,j_{12}) - 20 t_{\rm p})\,,
\eea
where $\Theta$ denotes the step function. The binding energy $E = m_1 m_2 /2 r_a$ \eqref{eq:2body_E} is determined from the initial separation $x$ by Eq.~\eqref{eq:r_a}. It gives $x^4 \approx 0.6 (m_1+m_2) m_1 m_2 / (\rho_R E)$. 

Consider now the integrals over $x$, $y$ and $\theta$ in \eqref{eq:niE0}. The $x$ integral can be eliminated using the $\delta$-function. The $y$ integral can be replaced with the integral over $\bar N(y)$ and the condition~\eqref{eq:taubound} can now be recast as an upper bound on $\bar N(y)$,
\bea
    \bar N(y) 
    &\leq \bar N(y)_{\rm max} \\
    &\equiv 3800 f_{\rm PBH} E^{-\frac{37}{36}} \rho_R^{\frac{1}{4}}  \frac{m_1^{\frac{11}{12}} m_2^{\frac{11}{12}} m_3^{\frac{7}{9}}(m_1+m_2+m_3)^{\frac{2}{9}}}{(m_1+m_2)^{\frac{11}{36}}\langle m\rangle} |\sin(2 \theta)|^{\frac{7}{9}}\,.
\eea
For consistency, we will further require that $\bar N(y)_{\rm max} < 1$ at every orientation, which yields a lower bound on the binding energy
\be\label{eq:E_min_E3}
    E 
    > E_{\rm min} 
    \equiv \langle m \rangle \left[11 {\rm \frac{km}{s}}\right]^2  \left[\frac{\langle m \rangle}{\Msun}\right]^{\frac{18}{37}} f_{\rm PBH}^{\frac{36}{37}}\mathcal{G}(m_1,m_2,m_3)
\ee
where we defined the dimensionless factor
\bea
    \mathcal{G}(m_1,m_2,m_3) 
    &= \left(\frac{2 m_1^3 m_2^3}{m_1+m_2}\right)^{\frac{11}{37}} m_3^{\frac{28}{37}} \left(\frac{m_1+m_2+m_3}{3}\right)^{\frac{8}{37}}\langle m\rangle^{-\frac{91}{37}}
\eea
that becomes unity for monochromatic mass functions. With this cut-off, we can simplify the computation and omit the exponential suppression in \eqref{eq:n_b3}, so the integral over $\bar N(y)$ can be approximated by $\bar N(y)_{\rm max}$.  

Averaging over $\theta$ then gives the initial binding energy distribution
\bea\label{eq:n_i_3E}
    \frac{\td n_i(E,m_i)}{\td E \td m_1 \td m_2 \td m_3 } 
    &= \frac{\td n_{i,\rm mono}(E)}{\td E}  \mathcal{F}(m_1,m_2,m_3) \prod^{3}_{i=1} \frac{\td n}{n \, \td m_i}\,,
\eea
where 
\be
    \frac{\td n_{i,\rm mono}(E)}{\td E} 
    = 4.0 \times 10^{10} E^{-\frac{25}{9}} f_{\rm PBH}^{3} \langle m\rangle^{\frac{16}{9}}  \rho_R^{\frac{3}{2}}\,
\ee
and we defined the dimensionless factor
\bea
    \mathcal{F} (m_1,m_2,m_3)
    &= m_1^{\frac{5}{3}} m_2^{\frac{5}{3}} m_3^{\frac{7}{9}}\left(\frac{m_1+m_2}{2}\right)^{\frac{4}{9}}\left(\frac{m_1+m_2+m_3}{3}\right)^{\frac{2}{9}}\langle m\rangle^{-\frac{43}{9}} \,
\eea
that reduces to unity for monochromatic mass functions.

To convert Eq.~\eqref{eq:n_i_3E} into the energy distribution of binary parameters after the 3rd body is ejected, we assume that a 3-body system with binding energy $E'$ yields a binary with binding energy $E$
\be\label{eq:K}
	\frac{\partial K(E | E')}{\partial E}
	= \frac{\alpha}{E'}  e^{-\alpha (E/E'-1)} \Theta(E-E') \,.
\ee
This approximates the behaviour observed in numerical simulations of PBH formation in the early universe for which $\alpha = 1$ gives a decent fit~\cite{Raidal:2018bbj,Vaskonen:2019jpv}. The limiting case in which the initial binding energy does not change corresponds to the limit $\alpha\to\infty$. The Heggie-Hills law states that hard binaries tend to get harder, while soft binaries get softer~\cite{1975MNRAS.173..729H,1975AJ.....80..809H}. Thus, as the early binaries tend to be hard, decreasing $\alpha$ will support more sizeable energy changes and can thus model the effect of subsequent collisions with other PBHs in PBH clusters. Since we work with the assumption of narrow mass functions, we do not consider the dependence on the masses of the three interacting PBHs in Eq.~\eqref{eq:K}.

Typically the lightest PBH will be ejected from the 3-body system, as it is more likely to be accelerated beyond the escape velocity. So, the binary will be typically composed of the two heavier PBHs. To model this, we will assume that the probability of ejecting a PBH from the system is exponentially suppressed by the relative mass $p_i(m_i) \propto \exp( - \lambda m_i/(m_1+m_2+m_3))$, where $\lambda \geq 0$ is a free parameter. $\lambda = 0$ corresponds to equal probability of ejection, while for $\lambda \to \infty$ the lightest PBH is always expelled. The exact shape of this function must be determined numerically. Given $m_1$, $m_2$, $m_3$, the binary constituent mass distribution is
\bea
    P_m(m'_1,m'_2|m_1,m_2,m_3) 
    &= p_1 \delta(m'_1 - m_2)\delta(m'_2 - m_3) \\
    &+ p_2 \delta(m'_1 - m_1)\delta(m'_2 - m_3) \\
    &+ p_3 \delta(m'_1 - m_1)\delta(m'_2 - m_2)\,.
\eea

The distribution of orbital parameters of the early binaries from the 3-body channel is
\bea \label{eq:evo_n}
    \frac{\td n_{\rm E3}(j,E,m'_1,m'_2)}{\td j \td E \td m'_1 \td m'_2} 
    = \frac{\td P(j)}{\td j} &\int \td E' \td m_1 \td m_2 \td m_3
    \frac{\partial K (E | E')}{\partial E} \\
    &\times \frac{\td n_{i}(E')}{\td E'\td m_1 \td m_2 \td m_3} P_m(m'_1,m'_2|m_1,m_2,m_3)\,,
\eea
with the angular momentum distribution given by Eq.~\eqref{eq:Pj_pert}. Since $\partial n_i(E)/\partial E \propto E^{-25/9}$, it is possible to evaluate the $E'$ integral
\bea \label{eq:evo_n}
    \frac{\td n_{\rm E3}(j,E,m_1,m_2)}{\td j \td E}  
    &= \frac{\td P(j)}{\td j} \frac{\td n_{i,\rm mono}(E)}{\td E} \mathcal{K}(\alpha) \bar{\mathcal{F}}(m'_i) \frac{\td n(m_1)}{n} \frac{\td n(m_2)}{n}\,, 
\eea
where 
\bea\label{eq:barF_E3}
    \bar{\mathcal{F}}(m_1,m_2) 
    \equiv \int \frac{\td n(m)}{n} \bigg[ & (p_1(m,m_1,m_2) + p_1(m_1,m,m_2)) \mathcal{F}(m,m_1,m_2)  \\
    &+ p_3(m_1,m_2,m) \mathcal{F}(m_1,m_2,m) \bigg]
\eea
accounts for the variation in masses and the overall factor
\be\label{eq:K_E3}
     \mathcal{K}(\alpha) 
     \equiv \alpha^{-25/9} e^{\alpha} \Gamma \left(25/9,\alpha \right)
\ee
contains the effect of hardening in binary-single PBH collisions. $\Gamma \left(a,b \right)$ denotes the incomplete gamma function. We will use $\alpha = 1$, which corresponds to $\mathcal{K}(1) = 4.0$. As was discussed above, accounting for hardening via later encounter would reduce $\alpha$ and increase the merger rate. However, as the effect of later encounters was estimated to be small~\cite{Franciolini:2022ewd}, we will not consider this contribution. The enhancing effect of hardening might, however, become more relevant in scenarios in which PBHs are initially clustered because the probability of binary-PBH collisions is expected to increase in such cases.

When integrating over masses in Eq.~\eqref{eq:barF_E3} we neglected the lower bound $E_{\rm min}$~\eqref{eq:E_min_E3} on the binding energy $E$. To estimate $E_{\rm min}$ for the binary after the 3rd PBH is ejected, we make the substitution in Eq.~\eqref{eq:E_min_E3},
\be
    \mathcal{G}(m_1,m_2,m_3) \to \bar{\mathcal{G}}(M) 
    = \left(\frac{M}{2 \langle m\rangle}\right)^{91/37}\, ,
\ee
where $M$ is the total mass of the binary. As the binary is more likely to consist of the heaviest PBHs, $\bar{\mathcal{G}}$ will give a stricter lower bound than what would follow from $\mathcal{G}$, i.e., $\bar{\mathcal{G}} \gtrsim \mathcal{G}$. This will in turn soften the resulting merger rate. However, we expect the effect to be small. As with $\mathcal{G}$, we have that $\bar{\mathcal{G}} = 1$ for monochromatic mass functions.

The merger rate implied by the distribution of binaries \eqref{eq:evo_n} is given by Eq.~\eqref{eq:Rprel}. As with  $R_{\rm E2}$ in Eq.~\eqref{dR_early_y}, we can assume that the binaries formed at the Big Bang, so that
\bea \label{eq:Rp}
    \frac{\td R_{\rm E3}}{\td \ln m_1 \td \ln m_2}
    \!&=\!\int \td j \, \td E \, \frac{\td n_{\rm E3}(j,E)}{\td j \td E \td \ln m_1 \td \ln m_2} \delta\left(t - \tau(j,E) \right) \\
    &=\!R_{\rm E3, mono}\!
    \left(\frac{M}{2\langle m \rangle}\right)^{\frac{179 \gamma }{259}-\frac{2122}{333}}\!\!\!
    (4\eta)^{-\frac{3 \gamma }{7}-1}
    \bar{\mathcal{F}}(m_1,m_2)
    \psi(m_1)\psi(m_2) ,
\eea
where the prefactor corresponds to the merger rate for a monochromatic mass function
\be\label{eq:Rp_mono}
    R_{\rm E3, mono}
    = \frac{7.9 \times 10^4}{{\rm Gpc}^{3} {\rm yr}} 
    f_{\rm PBH}^{\frac{144\gamma}{259}+\frac{47}{37}} 
    \left[\frac{t}{t_0}\right]^{\frac{\gamma}{7}-1} 
    \left[\frac{\langle m \rangle}{\Msun}\right]^{\frac{5\gamma-32}{37}}
    \frac{e^{3.2 (1-\gamma)}\gamma}{28/9-\gamma} \mathcal{K} \,.
\ee
This merger rate depends weakly on the exact form of the condition~\eqref{eq:taubound}. For instance, imposing $\tau_i > q t_p(y)$, we get $R_{\rm E3} \propto q^{\frac{16(28-9\gamma)}{2331}}$.

We stress that the three-body merger rate estimate~\eqref{eq:Rp} holds for relatively \emph{narrow mass functions}. A hierarchical decoupling, i.e., $a(t_{12}) \ll a(t_{\rm p})$ implies a relatively narrow mass function which guarantees that $m_1+m_2$ is of the same order as $m_3$. If $m_3 \gg m_1+m_2$, then the gravitational potential of the third body could dominate the "1+2" system before its decoupling from the Hubble flow. On the other hand, if $m_3 \ll m_1+m_2$, the third body may not be heavy enough to disrupt the initial binary. So, extending these estimates to wider mass functions requires considering the cases involving large mass ratios separately.

\newpage
\section{PBH binary formation in the late Universe}
\label{sec:late}

In the late Universe, PBH binaries can form dynamically through two- or three-body interactions in dense structures that have decoupled from the Hubble flow. Binary formation in two-body encounters happens through the emission of GWs while in a three-body system, one of the bodies carries away kinetic energy from the system leaving the other two in a bound state. 

We characterize the PBH number overdensities with an effective parameter $\delta_{\rm eff}>1$. Consider first a structure that includes $N$ PBHs. In such a structure, the average PBH number overdensity is $\delta_N = N/(V_N n_{\rm PBH}) > 1$ where $n_{\rm PBH} = \sum_N n_N N$ is the comoving average PBH number density and $V_N$ is the volume containing the $N$ PBHs. The PBH merger rate is $R_N = \delta_N^q (\sigma_{v,N}/\sigma_v)^b R_1$ where $q=2$ for the two-body and $q=3$ for the three-body channel, $\sigma_{v,N}$ is the velocity dispersion of PBHs in a halo with $N$ PBHs, $\sigma_v$ is a reference value for the velocity dispersion, $b=-11/7$ for the two-body and $b\in (-55,-47)/7$ for the three-body channel (as shown in the following) and $R_1$ is the merger rate assuming the background value for the PBH number density and the velocity dispersion $\sigma_v$. The total PBH merger rate is then obtained by summing over different size halos:
\be
    R = \sum_N n_N V_N R_N = \frac{\sum_N n_N N \delta_N^{q-1} (\sigma_{v,N}/\sigma_v)^b}{n_{\rm PBH}} R_1 \equiv \delta_{\rm eff}^{q-1} R_1 \,,
\ee
where in the last step we defined the effective PBH overdensity $\delta_{\rm eff}$ that includes also the effect of the differences in the velocity dispersions in different halos. We assume that the mass distribution of PBHs in DM haloes matches their overall mass distribution.

\subsection{Two-body encounters}

\begin{figure}[t]
\begin{center}
\includegraphics[width=0.98\textwidth]{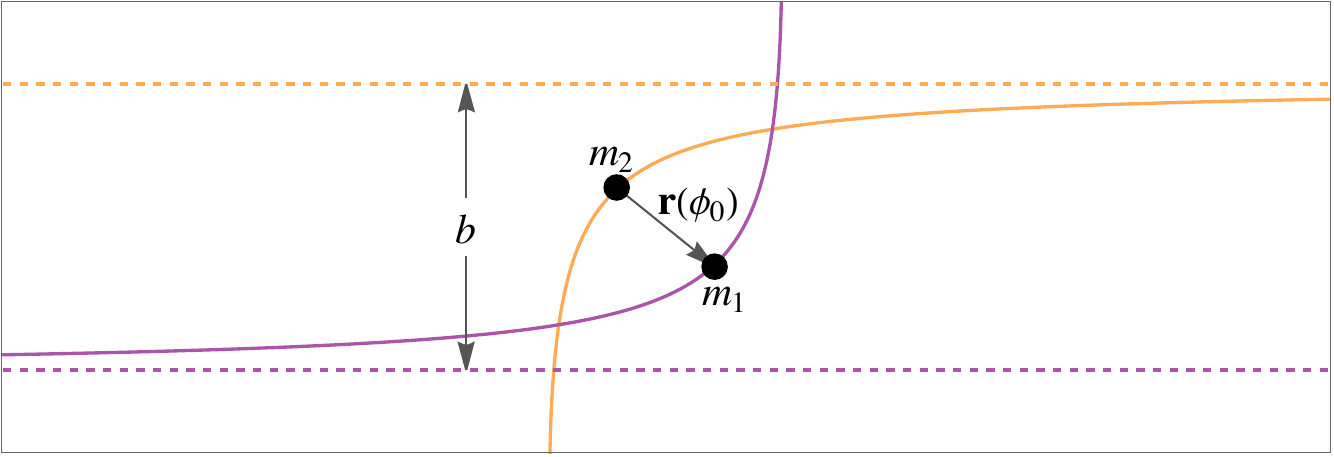}
\caption{A hyperbolic encounter of two PBHs.}
\label{fig:encounter}
\end{center}
\end{figure}

Consider a hyperbolic encounter of two bodies of masses $m_1$ and $m_2$ with initial relative velocity $v_{\rm rel}$ and impact parameter $b$, as shown in Fig.~\ref{fig:encounter}. In the centre of mass frame, the energy and angular momentum of the system can be written in terms of the polar coordinates of the separation $\vec r_1 - \vec r_2 \equiv (r\cos\phi,r\sin\phi,0)$ as
\be
    E = \frac12 \mu \left( \dot r^{2} + r^2 \dot\phi^2 \right) - \frac{\mu M}{r} \,, \qquad \vec{L} = \mu r^2 \dot\phi \hat{z} \,,
\ee
where $M = m_1 + m_2$ and $\mu = m_1 m_2/M$. Due to conservation of energy and angular momentum, the distance $r$ between the bodies evolves as
\be \label{eq:rhyper}
    r = \frac{b \sin\phi_0}{\cos(\phi-\phi_0) - \cos\phi_0} \,, \qquad \dot \phi = -\frac{b v_{\rm rel}}{r^2},
\ee
where $\phi_0 = \pi - \tan^{-1} (b v_{\rm rel}/ G M)$ corresponds to the distance of closest approach. 

The system loses energy by emitting GWs. The energy emitted by unit time is given by the third time derivative of the quadrupole moment tensor $Q_{ij} = m_1 r_{1,i} r_{1,j} + m_2 r_{2,i} r_{2,j}$ as $\dot E_{\rm GW} = \dddot Q_{ij} \dddot Q^{ij}/45$. By integrating over time and taking the limit of small $v_{\rm rel}$ we get
\be
	E_{\rm GW} = \frac{19\pi \eta^2 M^8}{6 b^7 v_{\rm rel}^7}   \,,
\ee
where $\eta = \mu/M$.

The bodies form a binary if the energy emitted in GWs is larger than the initial kinetic energy of the pair, $E_0 = \mu v_{\rm rel}^2/2$. This corresponds to the maximal impact parameter
\be
    b_{\rm max} = \left(\frac{19 \pi \eta}{3}\right)^{1/7} M v_{\rm rel}^{-9/7} \,.
\ee
The semimajor axis and eccentricity of the resulting binary are
\be
    r_{a,0} = -\frac{\eta M^2}{2 E_f} \,,\qquad e_0 = \sqrt{1+\frac{2E_f v_{\rm rel}^2 b^2}{\eta M^3}} \,,
\ee
where $E_f \equiv E_0 - E_{\rm GW}$, and by Eq.~\eqref{eq:tau_gen} the binary coalescence time is
\bea
    \tau &= \frac{19\pi M}{85 v_{\rm rel}^3} \sqrt{\frac{(b/b_{\rm max})^{21}}{1-(b/b_{\rm max})^{7}}} \\ 
    &\approx 1\,{\rm day} \frac{M}{\Msun} \left(\frac{v_{\rm rel}}{100\,{\rm km}/{\rm s}}\right)^{-3} \sqrt{\frac{(b/b_{\rm max})^{21}}{1-(b/b_{\rm max})^{7}}} \,.
\eea
This expression diverges for $b=b_{\rm max}$ since in that case the orbit is a parabola, but already for $b<0.99 b_{\rm max}$ the square-root factor is smaller than $4$ and the binaries merge very quickly.

The capture cross section is then given by $\sigma = \pi b_{\rm max}^2$. Assuming a Maxwell-Boltzmann velocity distribution $P(v) \propto v^{2} \exp\left(-v^{2}/\sigma_v^{2}\right)$, that matches reasonably well the results of N-body simulations~\cite{Mao:2012hf}, we get~\cite{1989ApJ...343..725Q,Mouri:2002mc,Bird:2016dcv}
\be \label{eq:vsigma}
	\langle v_{\rm rel}\sigma_2 \rangle = \int \!\td v_{\rm rel} \, v_{\rm rel} P(v_{\rm rel}) \pi b_{\rm max}^2 \approx 2\sqrt{\pi} \left( \frac{19\pi \,\eta}{3} \right)^{2/7} \frac{M^2}{\sigma_v^{11/7}} \,.
\ee
The velocity dispersion $\sigma_v$ typically takes a value in the $1-100\,{\rm km}/{\rm s}$ range. Since the coalescence times of the binaries formed through the two-body capture process are very short, we can directly translate the capture rate into the PBH merger rate:
\bea \label{eq:rateR2}
	&\frac{\td R_{\rm L2}}{\td \ln m_1 \td \ln m_2} = \langle v_{\rm rel}\sigma_{\rm mer} \rangle \delta_{\rm eff} \frac{\td n_{\rm PBH}}{\td \ln m_1} \frac{\td n_{\rm PBH}}{\td \ln m_2} \\
    &\hspace{5mm}= 2\sqrt{\pi} \left( \frac{19\pi \,\eta}{3} \right)^{2/7} \frac{M^2}{\sigma_v^{11/7}} \frac{\delta f_{\rm PBH}^2 \rho_{\rm DM}^2}{m_1 m_2}\psi(m_1) \psi(m_2) \\
    &\hspace{5mm}\approx \frac{3.4\times 10^{-6}}{{\rm Gpc}^3 {\rm yr}} f_{\rm PBH}^2 \,\delta_{\rm eff} \left(\frac{\sigma_v}{{\rm km}/{\rm s}}\right)^{-\frac{11}{7}} \eta^{-\frac57} \psi(m_1) \psi(m_2) \,.
\eea
Clustered PBHs tend to have also a smaller velocity dispersion which further enhances the rate~\eqref{eq:rateR2}.  Moreover, clustering tends to expel light PBHs from the regions of large $\delta_{\rm eff}$ which effectively causes spatial dependence of the PBH mass function $\psi(m)$.

\subsection{Three-body encounters}

The merger rate of binaries formed from in three-body encounters is estimated by multiplying the rate of two-body encounters by the probability that a third body happens to be close to the pair and estimating the coalescence times of these binaries~\cite{1976A&A....53..259A,Ivanova:2005mi,Ivanova:2010ia,2015ApJ...800....9M,Korol:2019jud,Kritos:2020wcl,Rodriguez:2021qhl,Franciolini:2022ewd}.

Using \eqref{eq:rhyper} at $\phi=\phi_0$, we find that in a hyperbolic encounter, the distance between the two bodies gets smaller than $r_a$ if the impact parameter is $b^2 < r_a^2 (1 + 2 M/(r_a v_{\rm rel}^2)) \equiv b_a^2$. Assuming a Maxwell-Boltzmann velocity distribution, the velocity averaged cross section for the PBHs to reach within distance $r_a$ is
\be
    \langle v_{\rm rel} \sigma_2 \rangle = \int \!\td v_{\rm rel} \, v_{\rm rel} P(v_{\rm rel}) \pi b_a^2 \approx 2\sqrt{\pi} r_a \sigma_v \left( 1 + \frac{2 M}{r_a \sigma_v^2} \right) \,.
\ee

The time the PBHs spend at a distance smaller than $r_a$ can be integrated from the solution \eqref{eq:rhyper}. In the limit of small $v_{\rm rel}$ this gives $\tau_a \approx \sqrt{8 r_a^3/(9 M)}$. Then, the probability that a third PBH passes the PBH pair within distance $r_a$ from its center of mass is
\bea
    P_3 &= \int \td v_3 \,P(v_3) \pi r_a^2 (v_3 \tau_a + r_a) \delta n_{\rm PBH} \\
    &= \left[\frac{4}{3} \sqrt{\frac{2\pi r_a^7}{M}} \sigma_v + \pi r_a^3 \right] \delta_{\rm eff} \,n_{\rm PBH}  \,.
\eea

The rate of three-body encounters can be approximated as $\propto P_3 \langle v_{\rm rel} \sigma_2 \rangle$. The encounter leads to the formation of a binary if the semimajor axis $r_a$ is sufficiently small, $r_a < r_{a,{\rm max}}$. The upper limit on $r_a$ is estimated by setting a lower limit on the binary hardness factor $\kappa$ defined by comparing the binary binding energy to the average kinetic energy of the surrounding PBHs:
\be
    \kappa \equiv \frac{E_{\rm bin}}{\langle m \rangle \sigma_v^2/2} = \frac{\mu M}{r_a \langle m \rangle \sigma_v^2} \,,
\ee
where $E_{\rm bin}$ is the binding energy given by Eq.~\eqref{eq:2body_E}. The threshold value of $\kappa > \kappa_{\rm min} = 5$ is often used in the literature (see e.g.~\cite{2015ApJ...800....9M,Rodriguez:2021qhl,Franciolini:2022ewd}). 

Unlike the two-body encounters, the three-body encounters produce wide binaries whose coalescence time can be even longer than the age of the Universe. To estimate the merger rate, we need to first estimate the angular momenta of the binaries. Analogously to the early universe 3-body scenarios \eqref{eq:Pj_pert}, we take
\be \label{eq:Pjgamma}
    \frac{\td P}{\td j} = \gamma j^{\gamma-1} \,,
\ee
where $\gamma \in [1,2]$. The value $\gamma=2$ corresponds to the thermal distribution~\cite{1919MNRAS..79..408J} but Refs.~\cite{2019Natur.576..406S,Raidal:2018bbj} find that the distribution could be superthermal with $\gamma < 2$.

With the above ingredients, the binary formation rate in the three-body encounters is 
\be
    \frac{\td R_{b}}{\td m_1 \td m_2 \td j \td r_a} = \frac{\td P}{\td j} \frac{\td P_3 \langle v_{\rm rel} \sigma_2 \rangle}{\td r_a} \Theta(r_a - r_{a,{\rm max}}) \delta_{\rm eff} \frac{\td n_{\rm PBH}}{\td m_1} \frac{\td n_{\rm PBH}}{\td m_2} \,.
\ee
Then, using Eq.~\eqref{eq:Rprel}, we get the merger rate of the binaries formed in the three-body encounters:
\bea
    &\frac{\td R_{\rm L3}}{\td \ln m_1 \td \ln m_2} 
    = \frac{7\pi}{24} f_{\rm PBH}^3 \delta_{\rm eff}^2 \left(\frac{1360 \eta \sigma_v^8 t}{3M}\right)^{\!\frac{\gamma}{7}} \frac{M^3 \rho_{\rm DM}^3}{\eta \langle m\rangle \sigma_v^9} \mathcal{F}\!\left(\frac{\langle m \rangle}{2 \eta M} \kappa_{\rm min}\right) \psi(m_1) \psi(m_2) \\
    &\approx \frac{1.3\times 10^{-16} e^{-6.0(\gamma-1)}}{{\rm Gpc}^3 {\rm yr}} f_{\rm PBH}^3 \delta_{\rm eff}^2 \left(\frac{\sigma_v}{{\rm km}/{\rm s}}\right)^{-9+\frac{8\gamma}{7}} \eta^{-1+\frac{\gamma}{7}} \left(\frac{M}{M_\odot}\right)^{3-\frac{\gamma}{7}} \left(\frac{t}{\tH}\right)^{\frac{\gamma}{7}} \\
    &\quad \times \mathcal{F}\!\left(\frac{\langle m \rangle}{2 \eta M} \kappa_{\rm min}\right) \psi(m_1) \psi(m_2) \,,
\eea
where 
\be \label{eq:Fkappa}
    \mathcal{F}(\kappa) \equiv \kappa^{-4+\frac{4\gamma}{7}} \left[ \frac{6\sqrt\pi}{7-\gamma}  + \frac{72}{63-8\gamma} \kappa^{-\frac12} + \frac{15 \sqrt\pi}{70-8\gamma} \kappa^{-1} + \frac{22}{77-8\gamma} \kappa^{-\frac32} \right] \,.
\ee

Similarly, as in the case of two-body encounters, the rate is larger in the higher density environments with smaller velocity dispersion. However, the three-body channel is even more sensitive to the clustering. Moreover, the coalescence time of the binaries formed through the three-body encounters is typically very large, so the binaries that would merge in the present Universe should have formed at a high redshift in compact PBH clusters induced by their Poisson distribution.

\newpage
\section{Summary}
\label{sec:summary}

In conclusion, the merger rate of PBH binaries consists of four components which we report here for standard PBH cosmologies:
\begin{enumerate}
    \item {\bf Early two-body channel} 
\be
\label{eq:1}
    \frac{\td R_{E2}}{\td \ln m_1 \td \ln m_2} \approx \frac{1.6 \times 10^{6}}{\Gpc^{3} \yr} \, f_{\rm PBH}^{\frac{53}{37}} \eta^{-\frac{34}{37}}  \left(\frac{M}{\Msun}\right)^{-\frac{32}{37}} \left(\frac{t}{\tH}\right)^{-\frac{34}{37}} S_{\rm L} S_{\rm E} \,\psi(m_1) \psi(m_2) \,,
\ee
where $S_{\rm L}$ and $S_{\rm E}$ are suppression factors given in Eqs.~\eqref{eq:S1_approx} and \eqref{eq:S2a}.

\vspace{1mm}
\item {\bf Early three-body channel} 
\bea
\label{eq:2}
    \frac{\td R_{\rm E 3}}{\td \ln m_1 \td \ln m_2} 
    &=  \frac{7.9 \times 10^4}{{\rm Gpc}^{3} {\rm yr}} 
    f_{\rm PBH}^{\frac{144\gamma}{259}+\frac{47}{37}} 
    \left[\frac{t}{t_0}\right]^{\frac{\gamma}{7}-1} 
    \left[\frac{\langle m \rangle}{\Msun}\right]^{\frac{5\gamma-32}{37}}
    \frac{e^{3.2 (1-\gamma)}\gamma}{28/9-\gamma} \mathcal{K} \\
    &\times 
    \left(\frac{M}{2\langle m \rangle}\right)^{\frac{179 \gamma }{259}-\frac{2122}{333}}\!\!\!
    (4\eta)^{-\frac{3 \gamma }{7}-1}
    \bar{\mathcal{F}}(m_1,m_2)
    \psi(m_1)\psi(m_2)
    \, ,
\eea
where $\mathcal{K}$ and $\bar{\mathcal{F}}$ are given by Eqs.~(\ref{eq:K_E3},\ref{eq:barF_E3}) and account for 3-body dynamics. Numerical simulations suggest that $\mathcal{K} \approx 4.0$. The second line contains the effect of extended mass functions.

\vspace{1mm}
\item {\bf Late two-body channel}
\be 
\label{eq:3}
	\frac{\td R_{\rm L2}}{\td \ln m_1 \td \ln m_2} \approx \frac{3.4\times 10^{-6}}{{\rm Gpc}^3 {\rm yr}} f_{\rm PBH}^2 \,\delta_{\rm eff} \left(\frac{\sigma_v}{{\rm km}/{\rm s}}\right)^{-\frac{11}{7}} \eta^{-\frac57} \psi(m_1) \psi(m_2) \,,
\ee
where $\delta_{\rm eff}$ and $\sigma_v$ characterize the PBH density contrast and velocity dispersion in the structures.

\begin{figure}[t]
\centering
\includegraphics[width=0.94\textwidth]{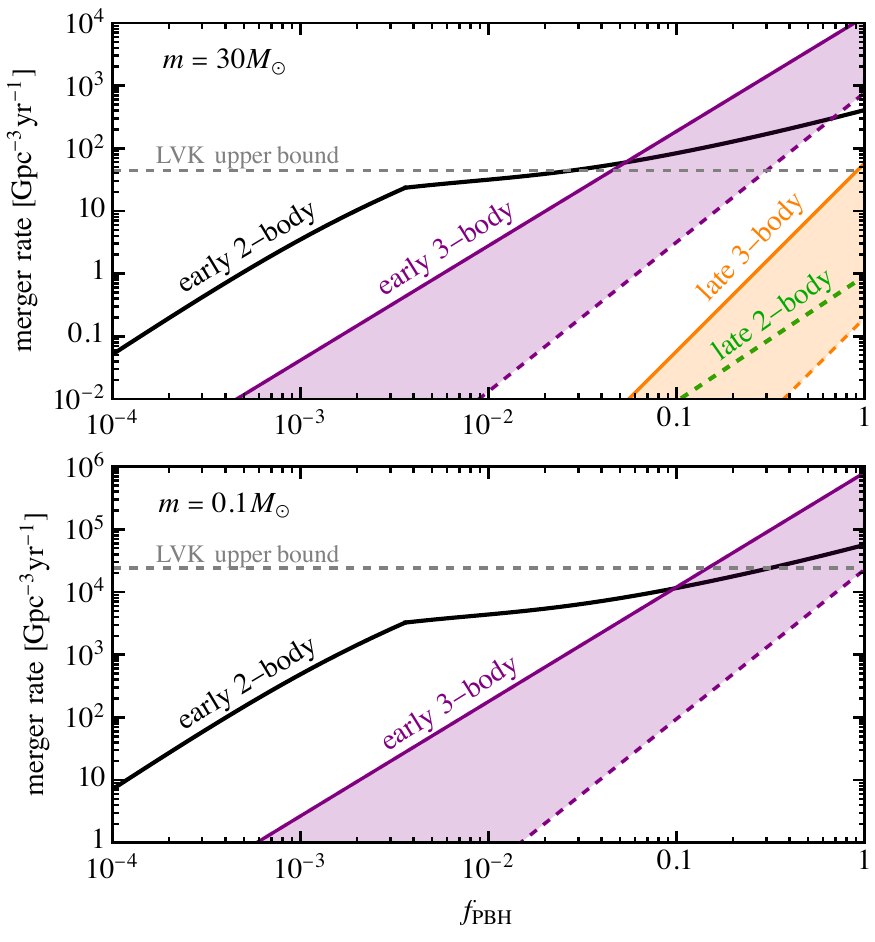}
\caption{Numerical estimates of the four channels contributing to the PBH merger rate as functions of the fraction of DM in PBHs. We assume a monochromatic mass function with $m=30\,\Msun$ (upper panel) and $m=0.1\,\Msun$ (lower panel). The bands show variations due to uncertainties in the angular momentum distribution. We show optimistic late universe merger rates with $\delta_{\rm eff} = 10^5$ for the late two-body and $\delta_{\rm eff} = 10^7$ for the late three-body channels. The gray dashed lines show the LVK upper bounds from~\cite{KAGRA:2021duu} and \cite{LIGOScientific:2021job}.}
\label{fig:mergerrates}
\end{figure}

\vspace{1mm}
\item {\bf Late three-body channel}
\bea
\label{eq:4}
    \frac{\td R_{\rm L3}}{\td \ln m_1 \td \ln m_2} \approx &\frac{1.3\times 10^{-16} e^{-6.0(\gamma-1)}}{{\rm Gpc}^3 {\rm yr}} f_{\rm PBH}^3 \delta_{\rm eff}^2 \left(\frac{\sigma_v}{{\rm km}/{\rm s}}\right)^{-9+\frac{8\gamma}{7}} \\
    &\times \eta^{-1+\frac{\gamma}{7}} \left(\frac{M}{M_\odot}\right)^{3-\frac{\gamma}{7}} \left(\frac{t}{\tH}\right)^{\frac{\gamma}{7}} \mathcal{F}\!\left(\frac{\langle m \rangle}{2 \eta M} \kappa_{\rm min}\right) \psi(m_1) \psi(m_2) \,,
\eea
where $\delta_{\rm eff}$ and $\sigma_v$ characterize the PBH density contrast and velocity dispersion in the structures, $\mathcal{F}(\kappa)$ is given in Eq.~\eqref{eq:Fkappa} and $\gamma$ determines the initial angular momentum distribution~\eqref{eq:Pjgamma} of the binaries.
\end{enumerate}

To demonstrate the relative importance of different PBH merger rate channels for different values of $f_{\rm PBH}$, we present numerical examples of the merger rates~\eqref{eq:1},~\eqref{eq:2},~\eqref{eq:3},~\eqref{eq:4} in Fig.~\ref{fig:mergerrates}. For concreteness, we have chosen a monochromatic mass function with $m = 30\,\Msun$. It is evident that for $f_{\rm PBH} \lesssim 0.1$ the early two-body channel~\eqref{eq:1} dominates. This channel is also best understood and has the smallest uncertainties. The situation may change when $f_{\rm PBH} \gtrsim 0.1$, that is, when a large fraction of DM of the Universe should be in the form of PBHs. In this case, the PHB density is large and the early three-body channel~\eqref{eq:2} may be the dominant one. However, the uncertainties associated with this solution are larger than in the first case. The upper (solid) and lower (dashed) boundaries of the three-body channels correspond to the merger rate with  $\gamma = 1$ and $\mathcal{K}=4$ and $\gamma=2$ and $\mathcal{K}=1$, respectively. The late Universe channels~\eqref{eq:3} and~\eqref{eq:4} are always expected to be subdominant. Optimistically, in Fig.~\ref{fig:mergerrates} we have fixed $\sigma_v = 1\,{\rm km}/{\rm s}$ and used $\delta_{\rm eff} = 10^5$ for the late two-body and $\delta_{\rm eff} = 10^7$ for the late three-body channel where we also show $\gamma = 1$ (upper solid) and $\gamma = 2$ (lower dashed) cases. Yet the late-time channels cannot be enhanced enough to be dominant. 

The current upper bound from GW observations by LIGO-Virgo-KAGRA (LVK) observational run O3 is indicated by the gray dashed lines in Fig.~\ref{fig:mergerrates}. Comparing it with the predicted merger rate will implies a constraint on $f_{\rm PBH}$. However, the observational bound on the merger rate is model dependent -- it is affected by the characteristics of the binary population, e.g., the mass and redshift dependencies in the merger rates. Thus a dedicated analysis is needed for accurate GW constraints on $f_{\rm PBH}$ and Fig.~\ref{fig:mergerrates} can at best indicate the order of magnitude. Given the present sensitivity to $f_{\rm PBH}$, both early channels can be relevant to observations. The early two-body channel constraints $f_{\rm PBH} < \mathcal{O}(10^{-3})$ for heavier $m \in 1-100\Msun$ PBHs, while the early three-body channel implies that the abundance sub-solar mass PBHs should not exceed $\mathcal{O}(10\%)$~\cite{Hutsi:2020sol, Franciolini:2022tfm, Andres-Carcasona:2024wqk}.

Initial clustering of PBHs affects\footnote{The effect of clustering will affect constraints on the PBHs. In the stellar mass range, which may be probed by terrestrial GW interferometers, it will strengthen constraints on the PBH abundance arising from the Lyman-$\alpha$ observations~\cite{DeLuca:2022uvz}. As a result, only relatively mild initial clustering is permitted in the mass range that is currently observational available.}, in particular, the early Universe binary formation channels. A rough estimate of the effect of initial clustering can be obtained by considering a constant 2-point function at the scales relevant for binary formation, $1+\xi(\vec{x}) = \delta$. As this is equivalent to a local change in the PBH number density, its effect on the merger rate is a simple rescaling: $R(f_{\rm PBH}) \to \delta^{-1} R(\delta f_{\rm PBH})$. However, this scaling does not hold for later processes, such as disruption in haloes, that do not depend on the local PBH density at their formation. The merger rate of the early three-body channel is proportional to a higher power of $f_{\rm PBH}$ than that of the early two-body channel so it is also more sensitive to the initial clustering. Moreover, the initial clustering would enhance the disruption probability of the early binaries, suppressing $R_{E2}$. On the other hand, $R_{E3}$ gets enhanced even more as the three-body encounters are more likely.

\vspace{1cm}
{\footnotesize {\bf Acknowledgments.}
This work was supported by the European Regional Development Fund through the CoE program grant TK202 and by the Estonian Research Council grants PRG803, PSG869, RVTT3 and RVTT7. The work of V.V. was partially supported by the European Union's Horizon Europe research and innovation program under the Marie Sklodowska-Curie grant agreement No. 101065736.}

\newpage
\bibliographystyle{unsrt}
\bibliography{authorsample.bib}

\end{document}